\documentclass[acmtog,screen,nonacm]{acmart}
\acmSubmissionID{840}

\makeatletter
\@namedef{ver@everyshi.sty}{}
\makeatother

\AtBeginDocument{%
  }

\usepackage{my}
\graphicspath{{./smallimages/}}

\begin{document}

\title{\name: Generating Simple Quad Layouts via Chart Distance Fields}

\author{You-Kang Kong}
\authornote{Work done during internship at Microsoft.}
\email{kykdqs@gmail.com}   
\affiliation{%
  \institution{Tsinghua University}
  \city{Beijing}
  \country{P.~R.~China}
}
\author{Yang Liu}
\authornote{Corresponding author}
\email{yangliu@microsoft.com}  
\affiliation{%
  \institution{Microsoft Research Asia}
  \city{Beijing}
  \country{P.~R.~China}
}
\author{Yue Dong}
\email{yuedong@microsoft.com}  
\affiliation{%
  \institution{Microsoft Research Asia}
  \city{Beijing}
  \country{P.~R.~China}
}
\author{Xin Tong}
\email{xtong.gfx@gmail.com}  
\affiliation{%
  \institution{Microsoft Research Asia}
  \city{Beijing}
  \country{P.~R.~China}
}
\author{Heung-Yeung Shum}  
\email{msraharry@hotmail.com}
\affiliation{%
  \institution{Tsinghua University \& International Digital Economy Academy}
  \city{Beijing}
  \country{P.~R.~China}
}

\authorsaddresses{}

\begin{abstract}
  3D shapes from scanning, reconstruction, or AI-generated content often lack simple quad mesh layouts—critical for efficient editing and modeling. Existing quad-remeshing techniques typically produce complex layouts with irregular loops, leading to tedious manual cleanup and extensive algorithm tuning.
We introduce \name, a diffusion-based generative framework that leverages Chart Distance Fields (CDF) to synthesize simple quad layouts on 3D shapes. Our approach addresses two key challenges: (1) the discrete nature of mesh connectivity, which hinders learning, and (2) the scarcity of large-scale datasets with simple quad meshes.
To overcome the first, we propose CDF, a continuous surface-based representation enabling effective learning and synthesis of quad layouts. To address the second, we define loop-aware simplicity metrics and construct a large-scale dataset of high-quality quad layouts recovered from public 3D repositories through a robust quad-recovery pipeline.
Extensive evaluations across diverse 3D inputs show that \name consistently outperforms existing methods, producing robust, artist-friendly simple quad layouts.

\end{abstract}

\begin{teaserfigure}
    \centering
    \includegraphics[width=0.98\textwidth]{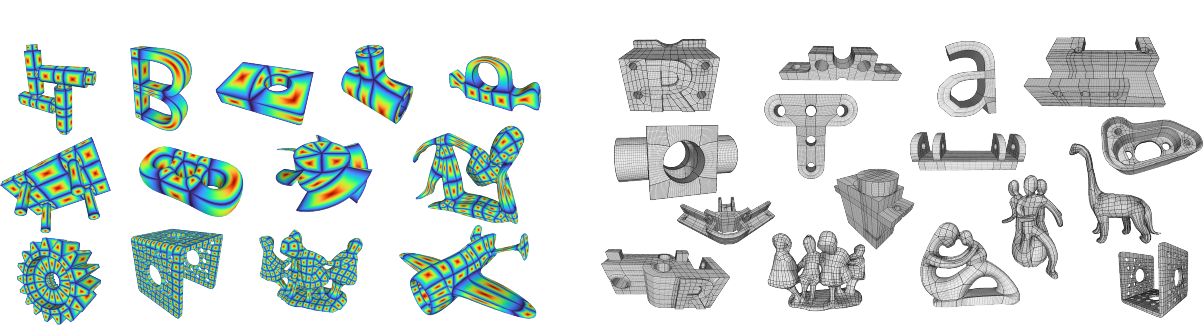}
    \Description{Teaser figure. SQuadGen synthesizes simple quad layouts on 3D shapes by learning to mimic quad layout patterns in chart distance field representation using a generative approach. Left: A gallery of synthesized CDFs across diverse 3D shapes.  Right: A gallery of quad meshes with simple layouts generated by SQuadGen. Complex chart boundaries are highlighted with thicker lines.}
    \caption{\name synthesizes simple quad layouts on 3D shapes by learning to mimic quad layout patterns in chart distance field representation using a generative approach. \textbf{Left}: A gallery of synthesized CDFs across diverse 3D shapes.  \textbf{Right}: A gallery of quad meshes with simple layouts generated by \name. Complex chart boundaries are highlighted with thicker lines.
    }
    \label{fig:teaser}
\end{teaserfigure}

\maketitle

\section{Introduction} \label{sec:intro}

Recent advances in AI-generated content (AIGC), along with widespread use of 3D scanning and reconstruction techniques, have introduced powerful tools for creating diverse and realistic 3D assets. However, the output formats --- such as triangle meshes, NeRF~\cite{nerf}, and 3D Gaussian Splatting~\cite{gaussiansplatting} --- rarely provide the \emph{simple quad mesh layouts} preferred by 3D artists. This limitation poses a significant barrier to integrating these 3D assets into established industrial modeling and editing pipelines.  \looseness=-1

By \emph{simple quad layout}, we refer to quad meshes that support standard loop-based modeling operations on face-loops, edge-loops, and edge-rings, with minimal irregular vertices. Such layouts exhibit fewer self-intersections and less spiraling, enabling intuitive region selection and loop-based editing. Moreover, well-structured quad layouts align naturally with shape geometry and feature lines, facilitating deformation, rigging, and simulation.

Existing quad-remeshing (retopology) techniques attempt to convert 3D shapes into quad meshes, but fully automatic tools often produce complex layouts with spiraling loops and poorly placed irregular vertices.
 This results in tedious manual cleanup and extensive algorithm tuning. Although recent methods~\cite{lyon2021simpler,campen2017partitioning} have improved layout quality, the input layout complexity still heavily influences the final result.

In this work, we present \name, a diffusion-based generative framework for synthesizing \emph{simple quad mesh layouts} optimized for editability and artistic workflows. Our approach addresses two key challenges: (1) \emph{How to represent discrete quad layout structures for effective learning?} and (2) \emph{How to curate a large-scale dataset of simple quad layouts for training?}

To tackle the first challenge, we introduce \emph{Chart Distance Fields} (CDF), a novel representation that transforms discrete quad layout structures into continuous surface fields. Together with its dual --- \emph{Dual Chart Distance Fields} (DCDF), CDF faithfully encodes quad layout structures and reframes quad layout generation as a continuous field synthesis problem, bypassing the complexity of directly predicting discrete mesh connectivity. Building on this representation, we develop a geometry-conditioned latent diffusion framework (\name) to synthesize CDFs across diverse 3D shapes and extract simple quad layouts.

For the second challenge, we design a loop-aware quad recovery algorithm that extracts quad layouts from large-scale 3D datasets such as Objaverse~\cite{deitke2023objaverse}, which primarily store triangle meshes originally derived from quad-dominant designs. Because layout quality varies, we introduce loop-aware metrics focusing on face-loop and edge-loop simplicity --- critical factors for editability. These metrics enable us to curate a dataset of over \SI{230}{k} simple quad layouts with diverse geometry. 

Trained on our curated dataset, \name achieves substantial improvements over popular quad-remeshing methods, consistently producing simple, artist-friendly layouts that facilitate efficient 3D editing and modeling while demonstrating strong generalization across diverse input geometries. 

To summarize, our contributions include:
\begin{enumerate}[leftmargin=*]\setlength\itemsep{1mm}
  \item[-] Introducing the \emph{chart distance field}, enabling simple quad layout generation as a continuous field synthesis task.
  \item[-] Developing a geometry-conditioned latent diffusion framework for generating simple quad layouts.
  \item[-] Curating a large-scale dataset of simple quad mesh layouts and proposing quantitative metrics for evaluating loop simplicity.
\end{enumerate}
We will release our code and model at \url{https://youkang-kong.github.io/squadgen} to facilitate future research.

\section{Related Work} \label{sec:relatedwork}

\paragraph{Quad Remeshing}
Quadrilateral remeshing of manifold surfaces is a longstanding problem~\cite{bommes2013quad}. Approaches include directional/cross-field design~\cite{ray2006periodic,vaxman2016directional,diamanti2014designing,diamanti2015integrable,corman2025rectangular}, mixed-integer programming~\cite{kalberer2007quadcover,bommes2009mixed,bommes2013integer,huang2018quadriflow,fang2018quadrangulation}, global quantization~\cite{campen2015quantized,lyon2021quad,coudert2024quad,heistermann2023min}, and edge-collapse techniques~\cite{jakob2015instant}. Data-driven methods optimize or learn cross fields~\cite{dong2025neurcross,dielen2021learning,dong2025crossgen,liu2025neuframeq,yu2025neural}, and heavily rely on external field-driven meshing algorithms to achieve good quality. Most methods prioritize field smoothness~\cite{liang2025field}, feature preservation~\cite{pietroni2021reliable}, anisotropy~\cite{lyon2020cost}, and planarity~\cite{liu2011general}, while layout simplicity --- critical for editability --- is often overlooked.

\paragraph{Simple Quad Layouts}
Prior works aim to reduce singularities for simpler layouts~\cite{ebke2016interactively,feng2021q,marcias2015data,cherchi2016polycube,campen2012dual,campen2014dual}, using strategies such as PolyCube topology~\cite{cherchi2016polycube}, curvature-guided dual loops~\cite{campen2012dual,campen2014dual}, and graph optimization~\cite{tarini2011simple,razafindrazaka2015perfect,pietroni2016tracing,viertel2019coarse,couplet2021generation}. Other approaches optimize initial layouts~\cite{bommes2011global,razafindrazaka2017optimal,lyon2021simpler}. Surveys~\cite{bommes2013quad,campen2017partitioning} provide comprehensive overviews. However, these methods depend heavily on input cross-fields or initial layouts, unlike our learning-based approach which bypasses these requirements.

\paragraph{Generative Mesh Modeling}
Recent research explores generative models for artist-preferred polygonal topologies. PolyGen~\cite{nash2020polygen} introduces an autoregressive model predicting mesh vertices and faces; follow-up works~\cite{siddiqui2023meshgpt,chen2024meshanything,weng2024pivotmesh,chen2024meshanythingv2,tang2024edgerunner,chen2024meshxl,shen2024spacemesh,hao2024meshtron,liu2025mesh} improve tokenizer design, mesh encoding, scalability, and quality. Diffusion-based techniques target triangle soups~\cite{alliegro2023polydiff} and Brep models~\cite{xu2024brepgen}. However, the manifoldness remains a challenge for existing works; only SpaceMesh~\cite{shen2024spacemesh} enforces edge manifoldness for closed surfaces. Concurrent work QuadGPT~\cite{liu2025quadgpt} proposes an autoregressive model for quad-dominant meshes with reinforcement learning to improve edge loops. In contrast, our method reframes quad meshing as surface-field synthesis, eliminating explicit connectivity prediction and focusing on remeshing existing shapes rather than generating new geometry.

\section{Overview}
\label{sec:overview}

We structure the paper as follows. In \cref{sec:layout}, we introduce key terminology and propose a loop simplicity metric to quantify quad layout complexity. \cref{sec:cdf} presents \emph{Chart Distance Fields} (CDF), a continuous representation derived from discrete quad layouts. In \cref{sec:method}, we present our core framework, \name, a geometry-conditioned latent diffusion model for synthesizing CDFs on given surfaces.  
\cref{sec:datacreation} details our data curation pipeline for training \name. Finally, \cref{sec:results} reports experimental results, ablation studies, and discusses limitations.

\section{Loop Simplicity of Quad Layouts} \label{sec:layout}

\subsection{Definitions and Notations} \label{subsec:def}

Let $\mQ$ denote a manifold quadrilateral mesh, with $\bv$, $\bf$, and $\be$ representing a vertex, face, and edge, respectively.
A vertex $\bv$ is \emph{regular} if it has four incident edges when interior, or three when on the boundary; otherwise, it is \emph{irregular} (or \emph{singular}).

\begin{figure}[t]
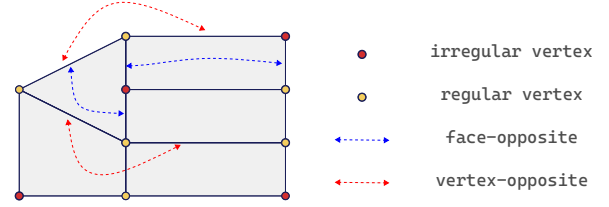

    \centering
    \imglabel{0.9\columnwidth}{term}{}
    \Description{Illustration of vertex regularity, vertex-opposite and face-opposite edge pairs.}
    \caption{Illustration of vertex regularity, vertex-opposite and face-opposite edge pairs.}
    \label{fig:term}
\end{figure}

At a regular vertex, incident edges form two pairs of \emph{vertex-opposite edges}: two edges that do not share a common face form one pair, and the remaining edges form the other. Each quadrilateral face has four edges, grouped into two pairs of \emph{face-opposite edges}.
\cref{fig:term} illustrates these concepts.

\begin{figure}[t]
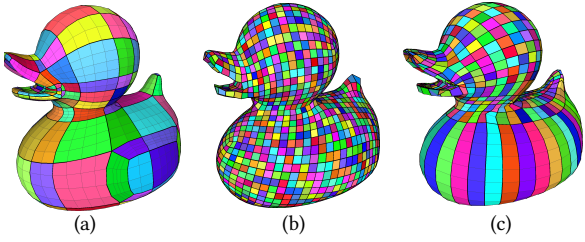

    \centering
    \imglabel{0.9\columnwidth}{complexillu}{%
        \put(0.12,-0.032){\small (a)}
        \put(0.48,-0.032){\small (b)}
        \put(0.84,-0.032){\small (c)}
    }
    \Description{
        Base complex illustration of different quadrilateral layouts. Each chart is colored randomly.
    }
    \caption{
        Base complex illustration of different quadrilateral layouts. Each chart is colored randomly.
    }
    \label{fig:complexillu}
    \vspace{-4mm}
\end{figure}

An \emph{edge-loop} is a polyline obtained by recursively traversing vertex-opposite edges starting from a given edge.
A \emph{separatrix} is a special edge-loop that starts at an irregular vertex and ends at another irregular or boundary vertex. All separatrices, together with mesh boundaries, partition the quad mesh into non-overlapping charts, forming the \emph{base complex}~\cite{campen2017partitioning}, denoted by $\mB$. To ensure each chart is homeomorphic to a quadrilateral patch, additional edge loops may be minimally inserted. The base complex encodes the quad mesh topology and thus represents the \emph{quad layout}. Its size, measured by the number of charts $N_c$, reflects structural simplicity: fewer charts imply simpler layouts and fewer singularities.
\cref{fig:complexillu} shows base complexes of three quad meshes, where (b) and (c) exhibit many charts, complicating editing.

Traversing face-opposite edges sequentially forms an \emph{edge-ring}, and the adjacent faces constitute a \emph{face-loop}. Edge-loops, face-loops, and edge-rings are widely used in 3D modeling software for mesh editing. Examples of edge-loops and face-loops appear in \cref{fig:loopillustration}, and the edge-ring associated with the face-loop in \cref{fig:loopa} is highlighted in yellow.

\subsection{Loop Simplicity} \label{subsec:metric}
Traditional indicators of quad layout simplicity, such as the number of base complex charts or irregular vertices, are coarse and fail to capture the complexity of loop-based editing. We introduce metrics based on \emph{face-loops} and \emph{edge-loops} to quantify layout quality in the context of quad mesh editing.

\paragraph{Face-loop and Edge-loop Metrics}
For a face-loop $\lf$ or edge-loop $\le$, we define two measures:
\begin{enumerate}[leftmargin=*]\setlength\itemsep{0mm}
  \item \textbf{Self-intersection count} ($\mSI$): number of repeated faces (or edge end-vertices) in the loop.
  \item \textbf{Rotation index} ($\Ind$): a measure of spirality. We form a polycurve by connecting face-loop centers or use the edge-loop polycurve, project it onto its best-fit plane, and compute total curvature divided by $2\pi$.
\end{enumerate}
\begin{definition}[Simple Loop]
  A loop is \emph{simple} if $\mSI = 0$ and $\Ind \leq 1$.
\end{definition}

Higher $\mSI$ and $\Ind$ indicate elongated, entangled regions that hinder loop-based editing and often require manual subloop selection.

\paragraph{Loop Simplicity}
To assess overall editability of a quad mesh, we compute the area ratio of regions controlled by simple loops:
\begin{equation}
  \bm{S}_{fl}(\mQ) := \frac{\sum_{\bf \in \mathcal{F}_s} \area(\bf)}{\sum_{\bf \in \mathcal{F}_{all}} \area(\bf)}, \quad
  \bm{S}_{el}(\mQ) := \frac{\sum_{\be \in \mathcal{E}_s} \area_f(\be)}{\sum_{\be \in \mathcal{E}_{all}} \area_f(\be)}.
\end{equation}
Here, $\mathcal{F}_{all}$ and $\mathcal{E}_{all}$ denote all face-loops and edge-loops, while $\mathcal{F}_s$ and $\mathcal{E}_s$ represent the subsets of simple loops. $\area(\bf)$ is the area of face $\bf$, and $\area_f(\be)$ is the total area of faces adjacent to edge $\be$. Both scores are bounded by \num{1}, achieved when all loops are simple. For example, grid or PolyCube-like layouts yield $\bm{S}_{fl} = \bm{S}_{el} = 1$ because they contain no spiral or self-intersecting loops.

We define the overall \emph{loop simplicity} as:
\begin{equation} \label{eq:loopsimplicity}
  \bm{S}_l(\mQ) = \min\bigl(\bm{S}_{fl}(\mQ), \bm{S}_{el}(\mQ)\bigr).
\end{equation}
Since edge-loops are typically shorter and subsets of face-loops, they tend to have lower rotation indices, making $\bm{S}_{el} \geq \bm{S}_{fl}$ in most cases. Thus, $\bm{S}_{fl}$ serves as a strong indicator. We use $\bm{S}_l$ in \cref{sec:datacreation} to filter training data for simple quad layouts.

\begin{figure}[t]
    \begin{subfigure}{0.44\columnwidth}
        \includegraphics[width=0.9\textwidth]
        {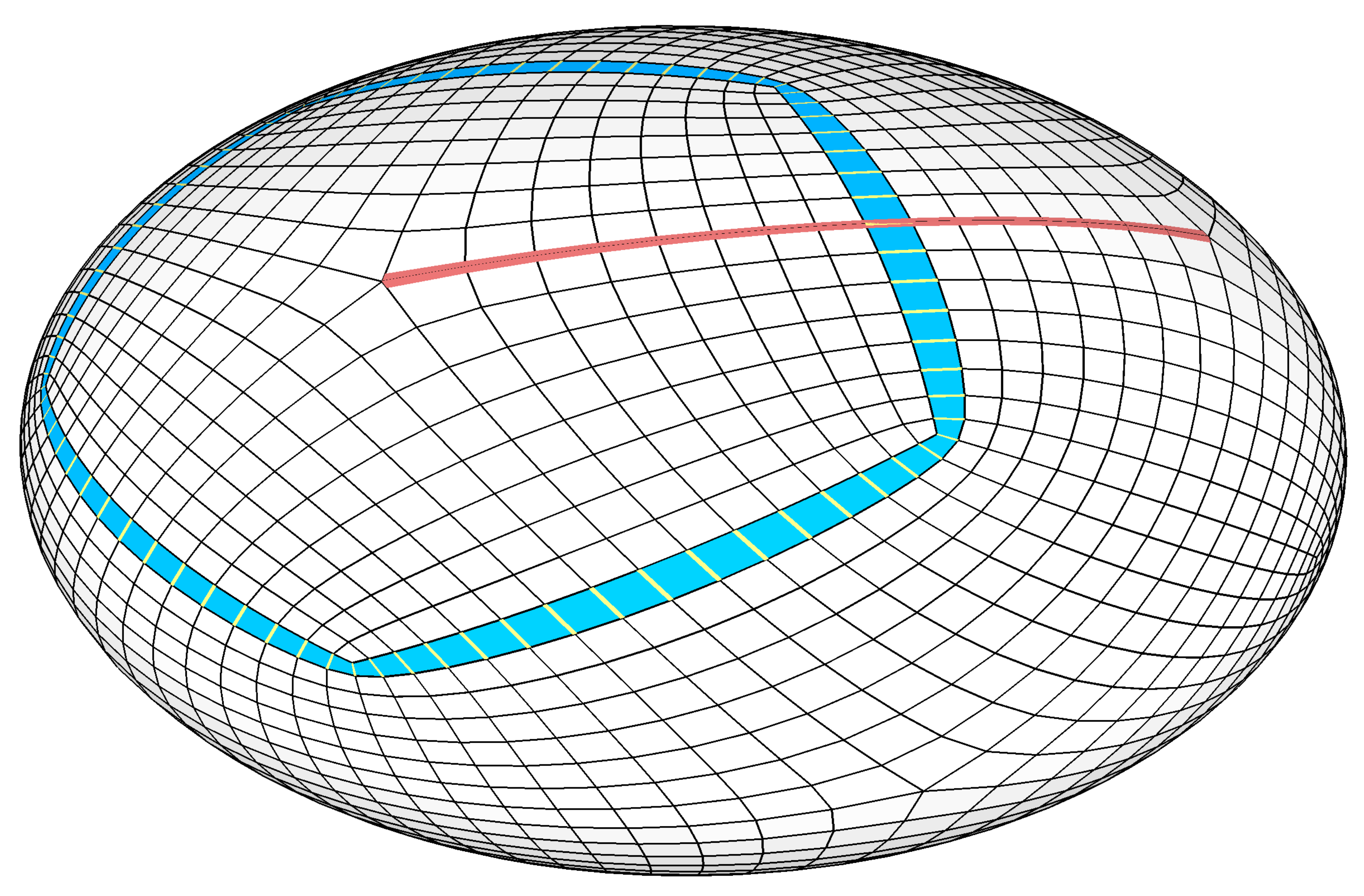}
        \caption{$\mathbf{S}_{fl} = \mathbf{S}_{el} = 1, N_c = 60$}
        \label{fig:loopa}
    \end{subfigure}
    \hfill
    \begin{subfigure}{0.44\columnwidth}
        \includegraphics[width=0.9\textwidth]
        {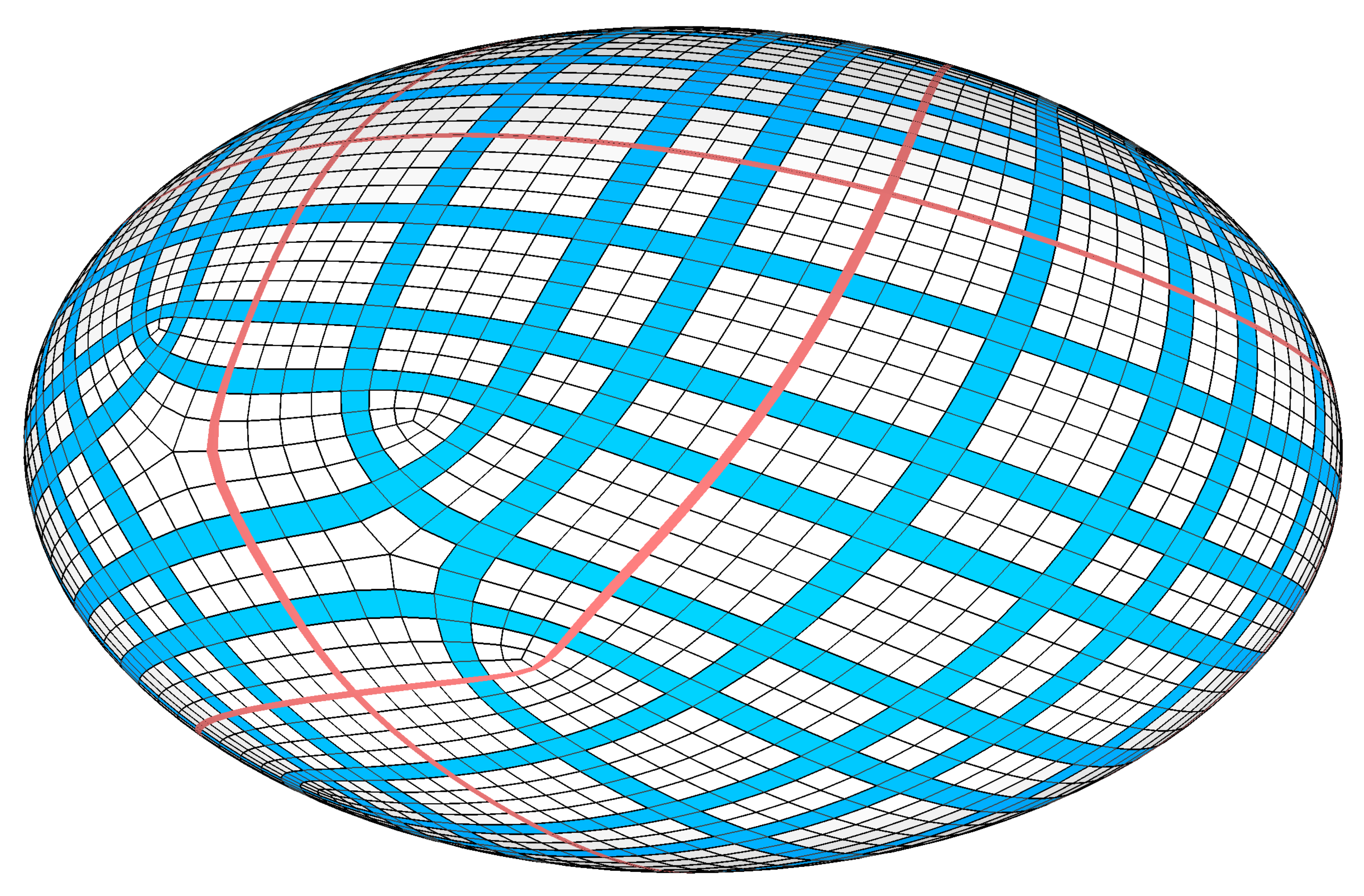}
        \caption{$\mathbf{S}_{fl} = 0, \mathbf{S}_{el} = 0.093, N_c = 436$}
        \label{fig:loopb}
    \end{subfigure}
    \Description{On each mesh, we select a face-loop (light blue) and an edge-loop (red) for illustration.  Loop simplicity scores and chart counts are reported. }
    \caption{On each mesh, we select a face-loop (light blue) and an edge-loop (red) for illustration.  Loop simplicity scores and chart counts are reported. }
    \label{fig:loopillustration}
    \vspace{-4mm}
\end{figure}

\cref{fig:loopillustration} shows examples: loops in \cref{fig:loopa} are all \emph{simple}, while those in \cref{fig:loopb} exhibit spirality and self-intersections, hindering editing.

\section{Chart Distance Field Representation}
\label{sec:cdf}

Learning and generating valid quad layouts is challenging due to their discrete, combinatorial nature. To address this, we introduce \emph{Chart Distance Fields} (CDF), a continuous representation of the base complex that enables effective learning of simple quad layouts.

CDF is conceptually intuitive: \emph{each chart in the base complex is treated as a curved quadrilateral face, and a normalized distance field is defined within the chart, decreasing smoothly from \num{1} at the chart center to \num{0} at its boundary along edge directions}. In the simplest case of a 2D unit square chart, the field reduces to the flipped $L_\infty$ distance function: $1-\max(|x|,|y|)$.

We further extend this idea to \emph{Dual Chart Distance Fields} (DCDF), defined analogously over dual charts formed by connecting neighboring chart centers.
In the following, we detail the computation of CDF and DCDF on a given quad mesh, and the connection to mesh parametrization and frame fields.

\subsection{Chart Splitting} \label{subsec:chartsplitting}
For a quad mesh $\mQ$, its base complex $\mB$ consists of charts $\{\mC_i\}$, each formed by $m_i \times n_i$ quad faces arranged in a grid-like pattern, where $m_i, n_i \in \mathbb{N}^+$. Each chart has four curved sides.

To define CDF, we first determine the chart center using an intuitive approach: \emph{connect the midpoints of opposite sides along edge-flow directions to form two flow lines; their intersection is the chart center}. These flow lines also split the chart into four subcharts, which simplifies subsequent CDF/DCDF computation.

The implementation proceeds as follows:
\begin{itemize}[leftmargin=*]\setlength\itemsep{1mm}
    \item \textbf{Edge length assignment}: For each edge-ring $\bm{r} \in \mQ$, compute its average edge length and assign this value to every edge $\be \in \bm{r}$. Under this metric, each chart $\mC_i$ can be treated as a rectangular grid patch. 
    \item \textbf{Chart splitting}: For each edge-loop inside a chart, locate its midpoint using the assigned lengths. Connecting midpoints of adjacent x- and y-direction edge-loops divides the chart into four disjoint subcharts; and chart center $\bm{c}_i$ which is shared by the four subcharts.
          Due to the edge length assignment, each subchart remains a quadrilateral patch.
    \item \textbf{Dual chart}: Subcharts adjacent to each chart corner collectively form a dual chart, whose center is defined at the chart corner. All dual charts collectively define the dual complex $\mB^\star$.
\end{itemize}

\begin{figure}[t]
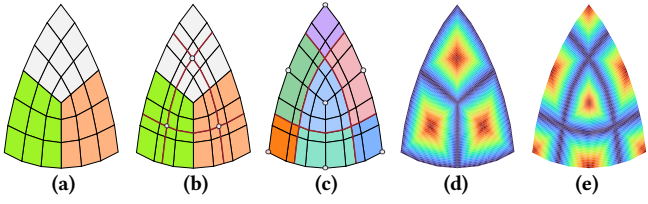

    \centering
    \imglabel{\columnwidth}{new2dillu}{
        \put(0.075,-0.03){\small \textbf{(a)}}
        \put(0.28,-0.03){\small \textbf{(b)}}
        \put(0.485,-0.03){\small \textbf{(c)}}
        \put(0.685,-0.03){\small \textbf{(d)}}
        \put(0.89,-0.03){\small \textbf{(e)}}
    }
    \Description{CDF and DCDF construction. (a) Input quad mesh with complex charts shown in different colors. (b) Chart splitting: red polylines indicate introduced edges, and circles mark chart centers. (c) Dual charts rendered in distinct colors, with circles marking dual chart centers. (d)\&(e) Colormaps of the CDF and DCDF. Dark red indicates 1, dark blue indicates 0.}
    \caption{CDF and DCDF construction.
        \textbf{(a)} Input quad mesh with complex charts shown in different colors.
        \textbf{(b)} Chart splitting: red polylines indicate introduced edges, and circles mark chart centers.
        \textbf{(c)} Dual charts rendered in distinct colors, with circles marking dual chart centers.
        \textbf{(d)\&(e)} Colormaps of the CDF and DCDF. Dark red indicates 1, dark blue indicates 0.
    }
    \label{fig:dispipeline}
\end{figure}

\cref{fig:dispipeline}(a-c) illustrates this process: red polylines connect midpoints, splitting original charts and producing dual charts.

\begin{figure}[t]
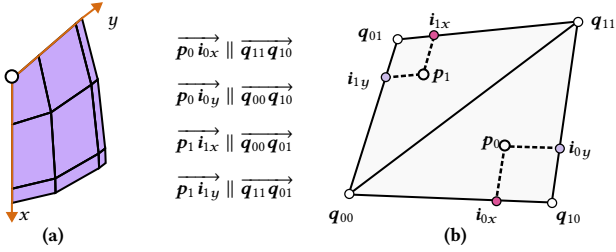

  \centering
  \imglabel{0.95\columnwidth}{coordnew}{}

  \Description{
    (a) Subchart coordinate system on a subchart. (b) Subchart coordinate computation at two sample points $\bm{p}_0$ and $\bm{p}_1$ inside a 3D quad face $\bm{q}_{00}\bm{q}_{10}\bm{q}_{11}\bm{q}_{01}$.
  }
  \caption{
    \textbf{(a)} Subchart coordinate system on a subchart.
    \textbf{(b)} Subchart coordinate computation at two sample points $\bm{p}_0$ and $\bm{p}_1$ inside a 3D quad face $\bm{q}_{00}\bm{q}_{10}\bm{q}_{11}\bm{q}_{01}$.
  }
  \label{fig:subchart}
  \vspace{-3mm}

\end{figure}

\subsection{Field Computation} \label{subsec:cdfcomputation}
Under the ring-based edge length metric, each subchart is treated as a rectangular grid. We define a \emph{subchart coordinate system} by placing the chart center at the origin and using two adjacent subchart corners as axis endpoints, aligned with edge-flow directions, as illustrated in \cref{fig:subchart}(a). These axes are normalized to unit length, and we assign \emph{subchart coordinates} to quad vertices within the subchart. These coordinates serve as scaffolds for computing CDF and DCDF.

For any point $\bm{p}$ inside a subchart, we compute its coordinates $(p_x,p_y)$ by projecting along edge directions rather than using barycentric interpolation. Specifically, we identify the quad containing $\bm{p}$, decompose it into two triangles, and emit rays from $\bm{p}$ parallel to x- and y-direction edges to find intersections with opposite edges.  \cref{fig:subchart}(b) illustrates this process at two sample points $\bm{p}_0$ and $\bm{p}_1$, and the resulting intersection points $\bm{i}_{0x}, \bm{i}_{0y}$ and $\bm{i}_{1x}, \bm{i}_{1y}$. These intersections allow interpolation along edge-flow lines, preserving alignment better than barycentric mapping. Detailed formulas for implementation are provided in \cref{appendix:subchart}.

\paragraph{Chart Distance Field}
Using subchart coordinates, the CDF for any point $\mp$ in a subchart $\mC_{sub}$ is:
\begin{align}
    \cdf(\mp) = 1 - \max(p_x, p_y), \quad \forall \mp \in \mC_{sub}.
    \label{eq:cdf}
\end{align}
Intuitively, the field value decreases smoothly from \num{1} at the chart center to \num{0} at its boundary.

\begin{figure}[t]
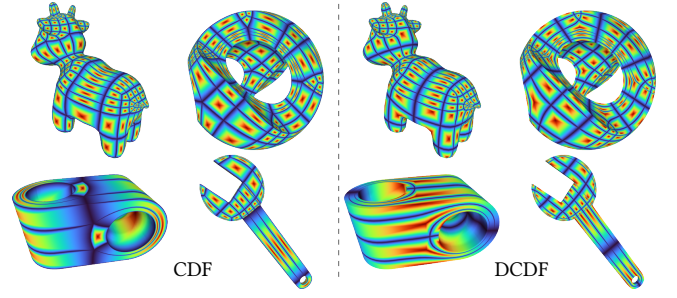

  \centering
  \imglabel{\columnwidth}{cdfvis}{
    \put(0.25,0.02){\small CDF}
    \put(0.75,0.02){\small DCDF}
  }
  \Description{CDF and DCDF visualization on four quad meshes.}
  \caption{CDF and DCDF visualization on four quad meshes.
  }
  \label{fig:dcdfvis}
  \vspace{-2mm}
\end{figure}

\paragraph{Dual Chart Distance Field}
Applying the same procedure to dual subcharts yields the DCDF. Since dual chart centers are opposite to chart centers, coordinates are complementary: $(p_x, p_y)$ becomes $(1 - p_x, 1 - p_y)$. Thus:
\begin{align}
    \dcdf(\mp) = 1 - \max(1 - p_x, 1 - p_y), \quad \forall \mp \in \mC_{sub}.
    \label{eq:dcdf}
\end{align}

By assembling fields across all subcharts, we obtain CDF and DCDF defined over the entire quad mesh, which form a periodical-like structure aligned with edge flows.

\cref{fig:dispipeline}(d,e) illustrates CDF and DCDF of the mesh in \cref{fig:dispipeline}(a).
Additional visualizations on more examples are provided in \cref{fig:dcdfvis}.

Both $\cdf$ and $\dcdf$ are $C^0$ continuous along quad edges, triangle edges introduced during quad decomposition, and points where $(p_x = p_y)$, and $C^1$ smooth elsewhere. Values range from \num{0} at chart boundaries to \num{1} at chart centers (CDF) or dual chart centers (DCDF).

\paragraph{Relation to Frame Fields and Global Parametrization}
At each subchart, the local coordinate system defines two axis directions aligned with edge flows, inducing a (generally non-orthogonal) frame field over the surface. The corresponding coordinates $(u, v) = (p_x, p_y)$ map each subchart to the unit square $[0,1]^2$, and together form a seamless global parametrization of the quad mesh.  However, recovering $(p_x,p_y)$ directly from CDF/DCDF introduces a branching ambiguity: solving \cref{eq:cdf,eq:dcdf} yields two possibilities,
\begin{equation}\label{eq:para}
    (u,v) = (\dcdf,\, 1-\cdf) \quad \text{or} \quad (u,v) = (1-\cdf,\, \dcdf),
\end{equation}
with the branch point occurring where $\cdf = \dcdf$ (the diagonal loci in \cref{fig:dispipeline}(d,e)).

\begin{figure}[t]
    \centering
    \includegraphics[width=\columnwidth]{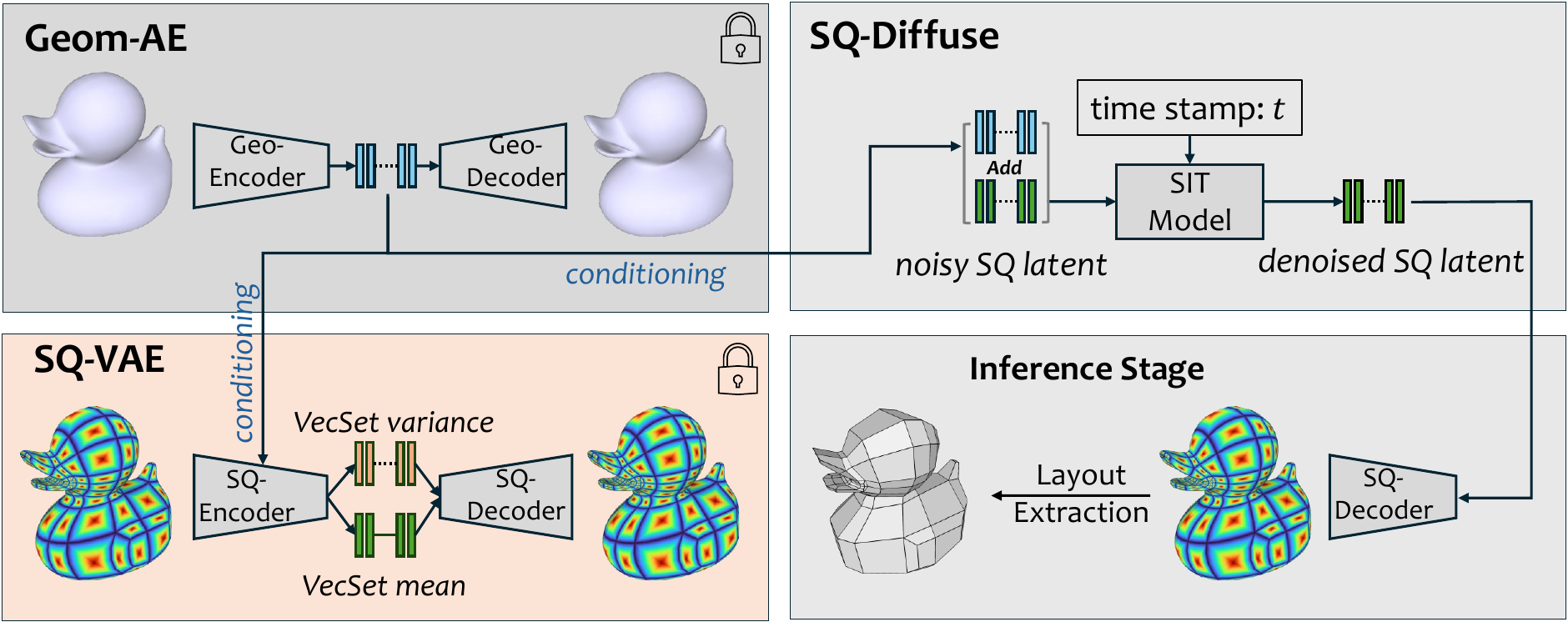}
    \Description{Design of \name. \name consists of three network components: \emph{Geom-AE} that encodes shape geometry; \emph{SQ-VAE} that learns a latent space for quad layouts; and \emph{SQ-Diffuse} that takes the geometry latent as conditions and denoises random noisy latent codes. The synthetic CDF is then converted to a quad layout via our layout extractor.  }
    \caption{Design of \name. \name consists of three network components: \emph{Geom-AE} that encodes shape geometry; \emph{SQ-VAE} that learns a latent space for quad layouts; and \emph{SQ-Diffuse} that takes the geometry latent as conditions and denoises random noisy latent codes. The synthetic CDF is then converted to a quad layout via our layout extractor.  }
    \label{fig:network}
    \vspace{-4mm}
\end{figure}
\begin{figure*}[b]
    \centering
    \includegraphics[width=\textwidth]{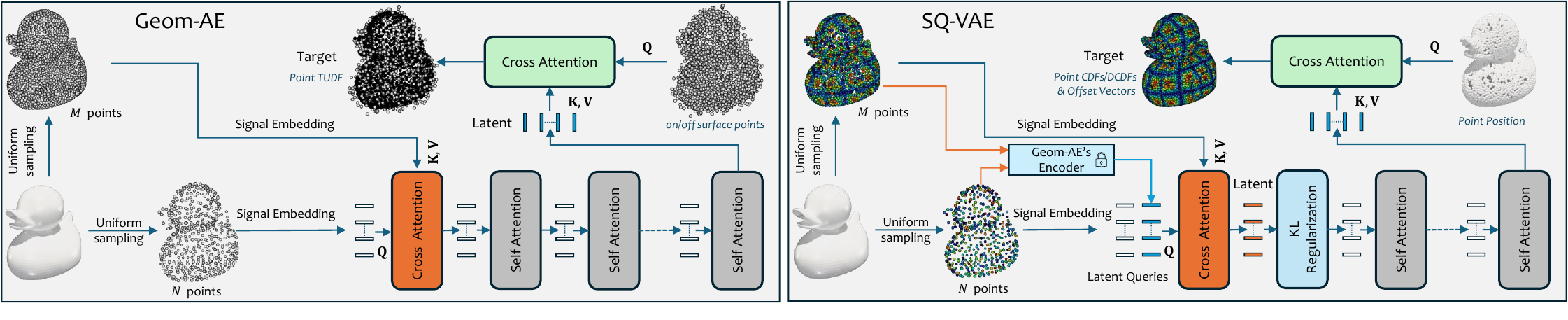}
    \Description{The network architecture of Geometry-AE and SQ-VAE. The architecture is based on the network design of 3DShape2VecSet.}
    \caption{The network architecture of Geometry-AE and SQ-VAE. The architecture is based on the network design of 3DShape2VecSet~\cite{zhang20233dshape2vecset}. }
    \label{fig:aevae}
\end{figure*}

\paragraph{Learning Perspective}
Learning global parametrizations directly is challenging due to discontinuities across chart boundaries. Similarly, learning frame fields~\cite{liu2025neuframeq,yu2025neural} typically relies on polyvector representations~\cite{diamanti2014designing}, which, although expressive, require integrability enforcement and suffer from scale sensitivity.
In contrast, our CDF/DCDF formulation acts as a \emph{structural surrogate} for frame fields: the scalar fields encode chart centers, boundaries, and flow directions implicitly through smooth level sets, without explicitly modeling frame orientations. This yields a continuous signal that is significantly easier to learn and naturally supports robust quad layout extraction.

\paragraph{Relation to Spectral Quadrangulation}
Because CDF/DCDF are periodic scalar fields, they bear conceptual similarity to spectral quadrangulation methods~\cite{dong2006spectral,huang2008spectral,Zhang2010wavebased,ling2015spectral}, which construct periodic scalar fields from Laplace-Beltrami eigenfunctions to guide quad extraction. However, unlike spectral methods whose scalar fields are constrained by geometry-dependent eigenfunctions, our CDF/DCDF fields are constructed directly from known quad layouts. This removes reliance on spectral structures and enables learning layout-aware scalar fields optimized for simplicity.

\section{Simple Quad Layout Generation} \label{sec:method}
We formulate simple quad layout generation as a scalar field synthesis problem by representing quad layouts as continuous fields (CDF and DCDF). Our framework, \name, consists of four main components:
\begin{enumerate}[leftmargin=*]\setlength\itemsep{1mm}
    \item \textbf{Geom-AE}: a VectorSet-based autoencoder that encodes shape geometry into a latent space.
    \item \textbf{SQ-VAE}: a geometry-conditioned VectorSet-based variational autoencoder that encodes simple quad layouts into a latent space.
    \item \textbf{SQ-Diffuse}: a geometry-conditioned latent diffusion model that synthesizes CDFs and DCDFs from input shape geometry.
    \item \textbf{Layout Extraction}: a robust algorithm that converts generated CDFs or DCDFs into simple quad layouts.
\end{enumerate}
The overall architecture of \name is illustrated in \cref{fig:network}. The following sections describe each component in detail.

\subsection{Geometry Autoencoder} \label{subsec:geomlatent}
\paragraph{AE Architecture}
Building on 3DShapeVecSet~\cite{zhang20233dshape2vecset}, we design a geometry autoencoder that encodes shape geometry into a compact latent space represented by a fixed-size set of latent vectors (VecSet). The network comprises \num{24} transformer blocks with \num{8} attention heads, each of dimension \num{64}. The latent space contains $N$ tokens, initialized by $N$ surface points sampled via Poisson disk sampling. An additional $M = 4N$ points are uniformly sampled from the surface and serve as \emph{Key} and \emph{Value} inputs for the first attention block. Each point is associated with its position and surface normal, which are used as initial features. We adopt a progressive training strategy~\cite{zhang2024clay} to increase $N$ from \num{512} to \num{4096}, improving network capacity. The network architecture is illustrated in \cref{fig:aevae}(left).

\paragraph{Loss Function}
Unlike the original VecSet approach, which decodes an occupancy field, we output a truncated unsigned distance field (TUDF) to handle shapes with open boundaries. During training, we sample \num{2048} points on the surface, \num{1024} near the surface, and \num{1024} within the bounding box. Their TUDF values are queried via the cross-attention module, and $L_1$ losses are computed with weights \num{1}, \num{1}, and \num{0.1}, respectively.

\paragraph{Geometry Latent}
We define the geometry latent as the output of the final attention block rather than the first cross-attention block (as in \cite{zhang20233dshape2vecset}). This choice ensures the latent captures sufficient geometric detail for conditioning SQ-VAE and SQ-Diffuse. The geometry decoder is implemented as the final cross-attention block, and each latent token has a feature dimension of \num{512}.
\subsection{SQ-VAE} \label{subsec:latentvae}

\paragraph{VAE Architecture} We adopt the 3DShapeVecSet VAE~\cite{zhang20233dshape2vecset} again to encode quad layouts. 
The architecture uses the same transformer blocks as our Geom-AE, except that the latent token dimension is set to \num{32}, and a KL regularization block is added for VAE training.  The network architecture is illustrated in \cref{fig:aevae}(right).

\paragraph{Encoder Input} Similar to Geom-AE, the input quad mesh is sampled into two sets of surface points: $N$ points as \emph{query} and $M$ points as \emph{key} and \emph{value} for the VAE encoder. These points are also passed through our pretrained Geom-AE to obtain the geometry latent, which conditions SQ-VAE. Since each token corresponds to one query point, we combine its token feature with the corresponding point feature to initialize the query feature. Specifically, we project both the point feature and geometry latent token to 512 dimensions and sum them. This operation acts as a specialized \emph{positional embedding}, improving VAE learning by providing richer geometric context than using the original $N$ points alone.

\paragraph{Point Features} To encode the quad layout effectively, each input point is equipped with the following feature signals:
\begin{enumerate}[leftmargin=*]\setlength\itemsep{1mm}
    \item Point position $\mp$ and normal $\bm{n}$;
    \item $\dcdf(\mp)$, $\cdf(\mp)$, and the unit surface gradients of both fields, \ie $\frac{\nabla \dcdf(\mp) - (\nabla \dcdf(\mp) \cdot \bm{n}) \bm{n}}{\|\nabla \dcdf(\mp) - (\nabla \dcdf(\mp) \cdot \bm{n}) \bm{n}\|}$ and $\frac{\nabla \cdf(\mp) - (\nabla \cdf(\mp) \cdot \bm{n}) \bm{n}}{\|\nabla \cdf(\mp) - (\nabla \cdf(\mp) \cdot \bm{n}) \bm{n}\|}$;
    \item Offset vectors from $\mp$ to the centers of the chart and dual chart where $\mp$ resides.
\end{enumerate}
Here, although DCDF or CDF alone can characterize the quad layout alone, due to surface sampling and VAE capacity, we find that including both fields, as well as their gradients and offset vectors, helps improve layout learning and the later layout extraction process.

\paragraph{Decoder Output and Loss Function} Given a query point sampled on the input mesh, the decoder outputs its CDF, DCDF, and the offset vectors via the cross-attention module. During training, we randomly sample \num{8192} points and apply an $L_1$ loss on the decoded signals with weight of \num{1}. Additionally, we introduce a KL-divergence loss on the VecSet latents with a weight of \num{0.001}.

\subsection{SQ-Diffuse} \label{subsec:dcdf_diffusion}
Given a quad layout dataset, we first use SQ-VAE to encode layouts into SQ latents. These latents are then used to train a geometry-conditioned latent diffusion model that generates SQ latents from noise. For this, we adopt a variant of the state-of-the-art Scalable Interpolant Transformer (SiT)~\cite{ma2024sitexploringflowdiffusionbased} as the diffusion backbone, --- a velocity-based model with a linear interpolant and sampling via the optimal regularized diffusion coefficient. The network comprises \num{24} transformer blocks, each with \num{16} attention heads of dimension \num{64}. To improve training stability, we apply QK-Norm~\cite{dehghani2023scalingvisiontransformers22} before attention operations and replace QK-Norm with RMSNorm~\cite{zhang2019rootmeansquarelayer} in all transformer blocks. Conditioning is achieved by element-wise addition of the geometry latent and the noisy SQ latent, as illustrated in \cref{fig:network}.  For balance the training efficiency and GPU memory usage, we set the number of SQ latent tokens to \num{4096} during the training.

\paragraph{SQ-Diffuse inference} We found that using a large number of latents improves VAE reconstruction quality and latent diffusion in the inference stage, even though only up to \num{4096} latents were used during training. Therefore, during inference, we take a triangle mesh as input, sample \num{8192} uniform points as queries to obtain the geometry latent, and feed it to the latent diffusion model as conditions.  This test-time upsampling strategy significantly enhances the quality of generated CDF/DCDFs, especially for complex geometries.

\begin{figure}[t]
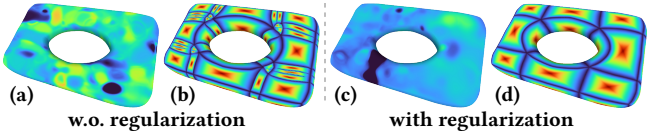

    \centering
    \imglabel{\columnwidth}{latentsmooth}{
        \put(0.01, 0.01){\small \textbf{(a)}}
        \put(0.26, 0.01){\small \textbf{(b)}}
        \put(0.515,0.01){\small \textbf{(c)}}
        \put(0.765, 0.01){\small \textbf{(d)}}
        \put(0.10,-0.03){\small \textbf{w.o. regularization}}
        \put(0.60,-0.03){\small \textbf{with regularization}}
    }
    \Description{Effect of regularized inference. Without regularization (left), the inferred quad layout (b), visualized as CDF, contains crowded, narrow patches and noticeable artifacts. With regularization (right), SQ-Diffuse generates a cleaner and more uniform layout (d). (a) and (c) visualize the initial CDFs from noise latents and their regularized version, respectively.}
    \caption{Effect of regularized inference. Without regularization (left), the inferred quad layout (b), visualized as CDF, contains crowded, narrow patches and noticeable artifacts. With regularization (right), SQ-Diffuse generates a cleaner and more uniform layout (d). (a)\&(c) visualize the initial CDFs from noise latents and their regularized version, respectively.}
    \label{fig:latentsmooth}
    \vspace{-2mm}
\end{figure}

\paragraph{Regularized inference} In our framework latent tokens are coupled with discretely sampled surface points. We find that regularizing their distribution based on 3D spatial proximity --- prior to the denoising process --- significantly enhances the structural coherence of the decoded output. This results in simpler, more regular quad layouts and improves generalization to out-of-distribution geometries. Specifically, we apply five iterations of a modified Taubin filter~\cite{taubin1995curve,taubin1995signal,taubin1996optimal} to the initial Gaussian noise. Each iteration performs a dual-pass smoothing over a $K$-nearest neighbor graph ($K = 32$):
\begin{align}
    \hat{\bm{z}}_i           & = \bm{z}_i + \lambda \sum_{j \in \mathcal{N}(i)} w_{ij} (\bm{z}_j - \bm{z}_i) / \sum_{j \in \mathcal{N}(i)} w_{ij},               \\
    \bm{z}_{i,\text{update}} & = \hat{\bm{z}}_i + \mu \sum_{j \in \mathcal{N}(i)} w_{ij} (\hat{\bm{z}}_j - \hat{\bm{z}}_i) / \sum_{j \in \mathcal{N}(i)} w_{ij},
\end{align}
where $\bm{z}_i$ denotes the latent vector at point $\bm{p}_i$, $\mathcal{N}(i)$ is the index set of its $K$ nearest neighbors, and $\lambda = 0.451$, $\mu = -0.472$ are smoothing parameters. The weight $w_{ij}$ is defined as:
\[
    w_{ij} = \exp\left(-\frac{\|\bm{p}_i - \bm{p}_j\|^2 + s \cdot\|\bm{n}_i - \bm{n}_j\|^2}{r \sigma_i^2}\right),
\]
where $\sigma_i$ is the minimal distance from $\bm{p}_i$ to its $K$ neighbors, and $\bm{n}$ denotes the point normal, $s=0.1, r=8$ are set in our experiments. The smoothed latent is then used as the initial input to the diffusion denoising process.

\cref{fig:latentsmooth} illustrates the impact of regularized inference, where the regularized noise latents yield a simpler layout.
We hypothesize that this regularization encourages the diffusion model to focus on generating low-frequency components of the latent distribution, thereby reducing noisy or irregular layouts.

\paragraph{Discussion}
The above regularization step effectively applies the Taubin operator
\(
\bm{A} = (\bm{I} + \mu \bm{L})(\bm{I} + \lambda \bm{L})^{k}
\)
linearly to the Gaussian noise latent $\epsilon \sim \mathcal{N}(0, \sigma^2 \bm{I})$, producing $z = \bm{A}\epsilon$. As a result, $z$ remains Gaussian, with an anisotropic covariance given by $\sigma^2 \bm{A}\bm{A}^\top$. Here, $\bm{L}$ denotes the graph Laplacian constructed on the $K$-NN graph, and $k = 5$ is the number of Taubin iterations we use. Ideally, the model should be trained using the same smoothed noise distribution so that the learned prior fully matches the test-time distribution. In our current setup, however, we find that applying mild Taubin smoothing only at inference --- without costly retraining and without significantly perturbing the original noise --- is sufficient for low-frequency CDFs. A more principled theoretical analysis, as well as a fully matched training-inference formulation, is left for future work.

\subsection{Quad Layout Extraction} \label{subsec:quad_extraction}

\begin{figure}[t]
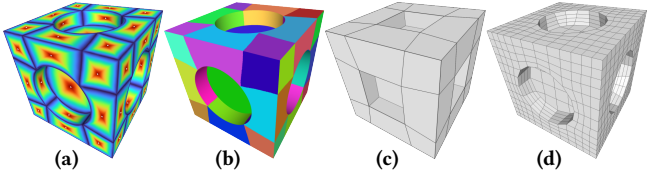

  \centering
  \imglabel{\columnwidth}{quadextraction}{
    \put(0.08,-0.01){\small \textbf{(a)}}
    \put(0.33,-0.01){\small \textbf{(b)}}
    \put(0.58,-0.01){\small \textbf{(c)}}
    \put(0.83,-0.01){\small \textbf{(d)}}
  }

  \Description{Layout extraction. (a) Input triangle mesh textured with the synthesized CDF. (b) Face clustering results, with clusters rendered in distinct colors. (c) Extracted layout mesh. (d) Refined mesh.}
  \caption{Layout extraction. \textbf{(a)}: Input triangle mesh textured with the synthesized CDF. 
  \textbf{(b)}: Face clustering results, with clusters rendered in distinct colors. \textbf{(c)}: Extracted layout mesh. \textbf{(d)}: Refined mesh.}
  \label{fig:extraction}
  \vspace{-2mm}
\end{figure}

Given a surface $\mathcal{S}$, our goal is to extract a quad layout from the synthesized SQ latent produced by SQ-Diffuse. The extraction pipeline consists of three main steps: discretization, face clustering, and layout extraction, followed optionally by refinement. \cref{fig:extraction} illustrates intermediate results.

\paragraph{Discretization} Assuming $\mathcal{S}$ is a triangle mesh, we first densify it via isotropic remeshing with sharp-feature preservation using PyMeshLab~\cite{pymeshlab}, typically yielding meshes with about half a million faces.
For each triangle center, we query its CDF values and offset vectors using the trained SQ-VAE decoder and synthesized SQ latents. Summing the offset vectors with face centers gives approximate chart and dual chart centers for each face $f$, denoted as $\bm{c}_{f, c}, \bm{c}_{f,dc}$.

\paragraph{Face Clustering}
Direct thresholding of CDF values is unreliable due to blur and gaps in the generated patterns. Instead, we adopt a robust extraction strategy that combines local maxima detection with priority-based region growing.
We first detect local maxima of CDF values on mesh faces, using an $r$-ring neighborhood for detection ($r=5$). These maxima serve as cluster seeds (\cref{fig:extraction}(a)). Faces with CDF values below $\delta = 0.1$ are marked and treated as ``walls''. Starting from each seed $s$, clusters grow by adding unmarked neighboring faces via a priority queue, where priority is determined by the proximity to the seed's estimated cluster center $\bm{c}_{s,c}$, measured as $|\bm{c}_{f,c} - \bm{c}_{s,c}|$. During region growing, we enforce manifoldness and, when possible, prevent propagation across sharp feature edges. After the initial growth phase, wall faces are unmarked and absorbed into adjacent clusters through continued region growing. We further merge disconnected clusters if more than \SI{50}{\percent} of their shared boundary faces have CDF values exceeding $0.5$.
The above parameters are chosen based on the triangle mesh density and the observed CDF/DCDF area distributions in our dataset. Within a reasonable range, these parameters are largely insensitive to the resulting layout combinatorics, and we therefore use a fixed configuration across all our experiments.
\cref{fig:extraction}(b) illustrates a representative clustering result.

\paragraph{Layout Extraction} Chart boundaries are obtained by traversing CDF cluster boundaries, forming a polygonal mesh. Edges corresponding to sharp features are tagged as \emph{feature edges}, and their vertices as \emph{feature vertices}. To simplify, we retain only patch corners as polygonal vertices. Since polygons may not be quadrilateral, we apply a feature-aware edge collapse procedure:
\begin{enumerate}[leftmargin=*]\setlength\itemsep{1mm}
    \item Assign collapse priorities by edge length (shortest first). Collapse edges if both adjacent faces are non-quads, the edge does not cross distinct sharp feature loops, and manifoldness is preserved. Feature vertex positions are retained.
    \item Repeat until no further collapses are possible.
\end{enumerate}
\cref{fig:extraction}(c) shows the extracted quad layout. We note that this step does not provide a theoretical guarantee of a pure-quad layout due to imperfections in the synthesized CDFs and the underlying geometry complexity; however, in practice, the vast majority of faces are quadrilateral.

\paragraph{Layout Refinement}
The extracted layout is refined into a dense quad mesh aligned with the input surface:
\begin{enumerate}[leftmargin=*]\setlength\itemsep{1mm}
    \item \textbf{Subdivision}: Apply midpoint subdivision to increase density while inheriting the sharp feature tags.  This step also converts quad-dominant meshes into all-quads.
    \item \textbf{Smoothing}: Improve mesh fairness using Winslow smoothing~\cite{knupp1999winslow}.
    \item \textbf{Projection}: Project refined vertices back onto the input mesh via nearest-point projection, ensuring feature vertices align with sharp feature lines.
\end{enumerate}
This process can be iterated for higher density (\cref{fig:extraction}(d)).

\paragraph{Implementation Note}
We represent quad layouts using a half-edge data structure. Certain configurations, \eg a cylinder with open ends decomposed into two patches whose boundary edges split into segments, cannot be represented directly in our current half-edge implementation, as the extracted layout would contain two quads sharing all four vertices. To resolve this issue, we utilize the parameterization implied by \cref{eq:para} and densify the CDF/DCDF patterns before layout extraction (\cref{appendix:densify} details the densification computation).
This effectively splits original patches into four smaller patches, resolving the topological issue -- with the following updated region-growing priority:  $\|\bm{c}_{f, c} - \bm{c}_{s, c}\| +\|\bm{c}_{f, dc} - \bm{c}_{s, dc}\|$.

\paragraph{Alternative Dual Chart Clustering} Dual chart clustering can be performed similarly using DCDF values and offset vectors. The layout extraction then follows the duality principle. However, imperfect DCDF patterns often produce non-manifold dual structures and need special handling. Therefore, we prefer CDF-based clustering for more robust quad layout extraction.

\section{Data Curation of Simple Quad Layouts} \label{sec:datacreation}

To enable learning-based generation of simple quad layouts optimized for easy editing, we curate a large-scale dataset of quad meshes with high loop simplicity. Our goal is not general quad remeshing --- existing tools applied to large-scale 3D datasets fail to produce layouts with the desired simplicity. Instead, we design a dedicated pipeline to collect and create high-quality quad layouts from public sources.
\subsection{Data from Triangle Merging} \label{subsec:objaverse}
\begin{figure}[t]
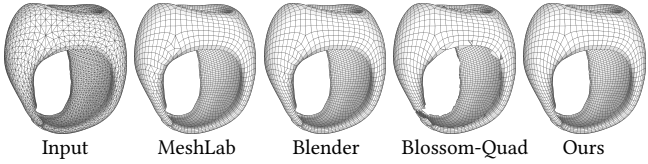

    \imglabel{\columnwidth}{tri2quad}{
        \put(0.06,-0.03){\small Input}
        \put(0.24,-0.03){\small MeshLab}
        \put(0.45,-0.03){\small Blender}
        \put(0.62,-0.03){\small Blossom-Quad}
        \put(0.87,-0.03){\small Ours}
    }
    \Description{Comparison of triangle-to-quad conversion algorithms. MeshLab, Blender, and Blossom-Quad leave 40, 56, and 94 triangles, respectively. The result from Blossom-Quad also exhibits non-manifold issues and unexpected holes.}
    \caption{Comparison of different triangle-to-quad algorithms. Comparison of triangle-to-quad conversion algorithms. MeshLab, Blender, and Blossom-Quad leave \num{40}, \num{56}, and \num{94} triangles, respectively. The result from Blossom-Quad also exhibits non-manifold issues and unexpected holes.}
    \label{fig:loopquad}
\end{figure}

Datasets such as Objaverse~\cite{deitke2023objaverse} and ABO~\cite{collins2022abo} contain large collections of 3D assets originally authored as quad-dominant meshes but stored as triangulated meshes, making direct recovery of quad layouts challenging. Naive triangle merging fails due to improper merge order and the Blossom-Quad algorithm ~\cite{remacle2012blossom} can reconstruct pure quads if the input derives from a quad mesh but struggles with open boundaries (see examples in \cref{fig:loopquad}).
To address these limitations, we propose a \emph{loop-quality-aware triangle merging algorithm} that prioritizes longer, direction-aligned face-loops while minimizing loop count. The algorithm consists of three steps.

\paragraph{Edge Sorting}
Sort all triangle edges by the \emph{rectangularity score} of the quad formed by adjacent triangles:
\[
    \text{Rectangularity} = \sum_{i=1}^4 |\angle_i - 90^\circ|,
\]
where $\angle_i$ are interior angles after projecting the quad to its normal plane. Edges are skipped if the dihedral angle is below \SI{120}{\degree} or the projected quad is non-convex.

\begin{figure}[t]
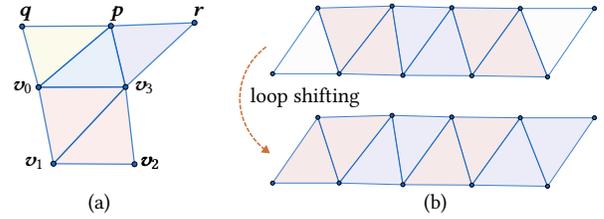

    \centering
    \imglabel{0.9\columnwidth}{loopmerge}{
        \put(0.01,0.04){\small $\bv_1$}
        \put(0.20,0.17){\small $\bv_3$}
        \put(-0.01,0.17){\small $\bv_0$}
        \put(0.21,0.04){\small $\bv_2$}
        \put(0.21,0.04){\small $\bv_2$}
        \put(0.16,0.30){\small $\bm{p}$}
        \put(0,0.30){\small $\bm{q}$}
        \put(0.30,0.30){\small $\bm{r}$}
        \put(0.12,-0.035){\small (a)}
        \put(0.70,-0.035){\small (b)}
        \put(0.40,0.15){\small loop shifting}
    }
    \Description{
        (a): Illustration of quad growing. Given a merged quad $\bv_0\bv_1\bv_2\bv_3$, the quad $\bv_0\bv_3\bm{p}\bm{q}$ merged via edge $\bv_0\bm{p}$ is more preferable than the quad $\bv_0\bv_3\bm{r}\bm{p}$ merged by edge $\bv_3\bm{p}$ due to its better alignment with the edge directions of $\bv_0\bv_1$ and $\bv_2\bv_3$. (b): Illustration of loop shifting. The two adjacent triangles in the same color corresponds to a merged quad.
    }
    \caption{\textbf{(a)}: Illustration of quad growing. Given a merged quad $\bv_0\bv_1\bv_2\bv_3$, the quad $\bv_0\bv_3\bm{p}\bm{q}$ merged via edge $\bv_0\bm{p}$ is more preferable than the quad $\bv_0\bv_3\bm{r}\bm{p}$ merged by edge $\bv_3\bm{p}$ due to its better alignment with the edge directions of $\bv_0\bv_1$ and $\bv_2\bv_3$. \textbf{(b)}: Illustration of loop shifting. The two adjacent triangles in the same color corresponds to a merged quad.}
    \label{fig:loopmerge}
    \vspace{-2mm}
\end{figure}

\paragraph{Direction-Aligned Loop Growing}
Process unmerged edges in descending rectangularity order. For each edge $\be$, initialize a priority queue and iteratively:
\begin{enumerate}[leftmargin=*]\setlength\itemsep{1mm}
    \item Merge the edge if possible, enqueue adjacent unmerged edges.
    \item For each merged quad, evaluate candidate merges using a \emph{misalignment score} --- the sum of angle deviations between opposite edges of the candidate and current quad. Lower scores indicate smoother edge-flow alignment. See \cref{fig:loopmerge}(a) for illustration.
\end{enumerate}

\paragraph{Loop Shifting}
Prioritizing rectangularity works well in most cases, but some artist-designed meshes contain parallelogram-like quads, which produce face-loops with dangling triangles (\cref{fig:loopmerge}(b)-top). To address this, we remerge face-loops along shared edges to absorb the triangles (\cref{fig:loopmerge}(b)-bottom), a process we call loop shifting. A loop shift is applied only if it (1) increases the number of quads, or (2) preserves the quad count while improving the alignment score. We apply this step recursively to all problematic face-loops until no further shifts are possible.

\cref{fig:loopquad} compares triangle merging algorithms implemented in Meshlab~\cite{cignoni2008meshlab}, Blender~\cite{blender}, and Gmsh's Blossom-Quad~\cite{remacle2012blossom} on a triangle mesh with open boundaries. Among these, only our method successfully recovers an all-quad mesh.

Our algorithm is still heuristic and does not guarantee recovery of pure quads even when the input mesh originates from a quad decomposition. In cases of failure, we fall back to the Blossom algorithm when it can succeed.

\paragraph{Data Processing} Meshes in Objaverse or other datasets typically contains multiple submeshes with non-manifold issues and mergeable seams.  We resolve these issues through boundary vertex merging  when possible; otherwise, the non-manifold submesh is discarded. We apply our triangle merging algorithm to convert triangles into quad-dominant meshes, retaining only pure-quad meshes.

\begin{figure*}[t]
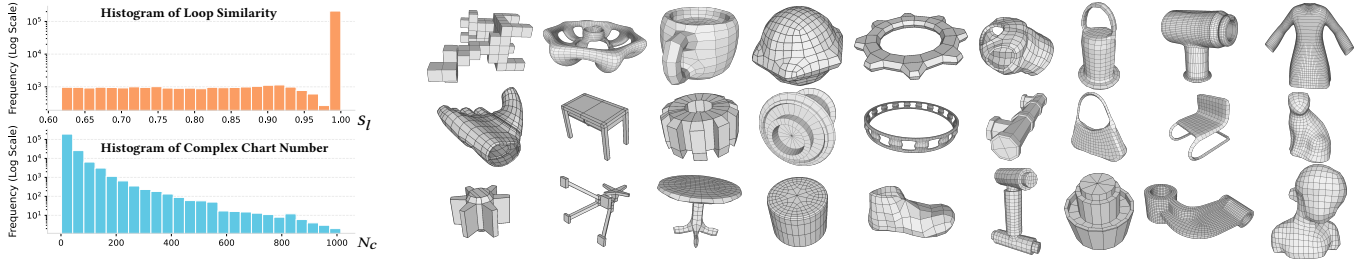

    \centering
    \imglabel{\linewidth}{datastat}{
        \put(0.07,0.18) {\tiny \textbf{Histogram of Loop Similarity}}
        \put(0.07,0.08) {\tiny \textbf{Histogram of Complex Chart Number}}
        \put(0.26,0.01){\tiny $N_c$}
        \put(0.26,0.10){\tiny $\bm{S}_l$}
    }
    \Description{Dataset statistics and visualization. Left: Histogram of loop similarity and complex chart number. Right: Quad meshes from our curated dataset.}
    \caption{Dataset statistics and visualization. \textbf{Left}: Histogram of loop similarity and complex chart number. \textbf{Right}: Quad meshes from our curated dataset.}
    \label{fig:datastat}
\end{figure*}
\subsection{Quad Meshes from Remeshing Tools}  \label{subsec:remesh}
Most artist-crafted models in Objaverse consist of numerous components with relatively simple geometry, our recovered quad meshes also exhibit simple structures. To enrich diversity, we collect quad meshes from public sources and generate new ones using state-of-the-art tools.

For triangle meshes from Objaverse, ABO~\cite{collins2022abo}, and ShapeNet~\cite{chang2015shapenet}, we first apply 3D Alpha Wrapping~\cite{portaneri2022alpha} to convert them into watertight triangle meshes with reduced geometric complexity. These preprocessed meshes are then retopologized using QuadRemesher~\cite{quadremesher}. In our experiments, QuadRemesher outperforms other automatic remeshing tools in terms of loop simplicity when the target quad count is constrained to the range $[300, 1000]$ and its symmetry option is enabled for shapes exhibiting symmetry.

Additionally, we incorporate quad meshes from prior research --- QuadWild~\cite{pietroni2021reliable}, and include boundary meshes of all-hexahedral meshes from previous all-hex meshing works, collected via the \href{https://www.hexalab.net/}{HexaLab} platform~\cite{bracci2019hexalab}.

\subsection{Data Curation} \label{subsec:datacuration}
We treat each single-connected component of collected models as an \textbf{individual quad mesh}, and normalize it to fit within a $[-1,1]^3$ bounding box after PCA alignment. We eliminate duplicates and near-duplicates by comparing geometric and mesh connectivity similarity across all meshes, resulting in \num{1.6} million single-connected quad meshes.
To ensure high loop simplicity and suitable layout complexity for training, we apply the following filtering criteria on these meshes:
\begin{enumerate}[leftmargin=*]\setlength\itemsep{1mm}
    \item \textbf{Loop simplicity score}: Discard quad meshes with a simplicity score below \num{0.618}; we choose this golden ratio to balance loop simplicity and geometric complexity: \emph{higher simplicity often correlates with simple geometry}.
    \item \textbf{Quad distortion}: Remove meshes with highly non-planar quads. 
    \item \textbf{Chart count}: Remove meshes with more than \num{1024} charts.
    \item \textbf{Chart area}: Exclude meshes containing any chart with area smaller than $1/1024$.
    \item \textbf{Chart side length}: Discard meshes with any chart side shorter than $\sqrt{1/1024}$.
\end{enumerate}
The last three criteria eliminate meshes with extremely small or narrow charts that are difficult to learn due to limited sampling resolution.  Additionally, we remove overly simple layouts with no singularities in the interior region (for open meshes) and meshes with excessive boundaries (more than 8).

The quality filtering reduces the dataset size to \textbf{\num{230}{k}}. The major contributions come from our triangle merging on Objaverse (\num{194}{k}) and QuadRemesher's results (\num{21}{k}).  \cref{fig:datastat} presents dataset statistics and visualizations of selected quad meshes from our curated dataset.

\section{Experiments and Evaluations} \label{sec:results}

\subsection{Experimental Setup} \label{subsec:setup}
We implemented and trained \name using PyTorch on 16 NVIDIA A100 GPUs with our curated dataset.
The Geom-AE, SQ-VAE, and SQ-Diffuse networks contain \SI{107}{M}, \SI{112}{M}, and \SI{802}{M} parameters, respectively.
All networks were trained using the AdamW optimizer. For Geom-AE and SQ-VAE, the learning rate was decayed from \num{1e-4} to \num{1e-6} following a cosine schedule, while for SQ-Diffuse it was kept constant at \num{1e-4}. Training both the VAE and diffusion models took approximately 14 days each.

\paragraph{Geom-AE Training} Since Geom-AE is unrelated to quad layouts, we directly use triangle meshes from Objaverse, ABO, and ShapeNet.  For models with multiple components, each component is treated as a single mesh. Additionally, we apply 3D alpha wrapping to convert models into single-component meshes. In total, \num{2.7} million meshes were prepared.  All meshes are normalized into a unit bounding sphere and augmented with random rotations during training.

\paragraph{Sharp-Feature-Aware Quad Layouts} For our dataset, when computing base complexes, we identify edge loops where every edge is sharp (dihedral angle less than \SI{130}{\degree}) and treat them as additional separatrices. This ensures that all sharp edge loops are preserved in the resulting quad layout as chart boundaries, encouraging the learning to respect sharp features.

\paragraph{SQ-VAE and SQ-Diffuse Training} We choose all unfiltered quad layouts to pretrain SQ-VAE, then fix the encoder and fine-tune the decoder with the curated layouts. During training, we apply two augmentations to improve robustness:
(1) Add slight random noise perturbations to input mesh vertices and apply small-scale and rotation changes;
(2) Apply two rounds of Catmull-Clark subdivision to create smoother versions.
The augmented and original data are combined for training SQ-VAE and SQ-Diffuse. While for SQ-Diffuse training, we use only curated data, and we also balance the data such that high-chart-number layouts appear more frequently in each training batch, to encourage the model to see more complicated geometry-layout pairs.

\paragraph{Inference} As introduced in \cref{subsec:dcdf_diffusion}, we use \num{8192} latents with the regularized inference strategy by default. The average inference time is \num{60} seconds, with the main computational bottleneck being self-attention over a large number of latents. With \num{4096} latents, the inference time reduces to \num{30} seconds. Layout extraction takes around \num{1} second. For each input, we synthesize four results and select the one with the best loop simplicity score for evaluation.  In mesh extraction step, we restrict the mesh subdivision level to at most $3$ and ensure that the total face number does not exceed $20000$.

\paragraph{Compared Methods} We compare \name against four representative quad-meshing approaches: QuadriFlow~\cite{huang2018quadriflow}, QuadWild with the Bi-MDF solver~\cite{heistermann2023min}, FSCP~\cite{liang2025field}, and QuadRemesher~\cite{quadremesher}. The first three are widely used or state-of-the-art quad-meshing methods, whereas QuadRemesher is a commercial tool known for producing high-quality quad meshes. We use their default parameters for all experiments.
We do not include comparisons with layout optimization methods reviewed in \cref{sec:relatedwork} due to the lack of publicly available implementations.

\paragraph{Evaluation Metrics}
For each extracted layout, we assess three simplicity indicators: the loop simplicity score ($\bm{S}_l$), the chart count ($N_c$), and the irregular vertex count ($N_I$). We also evaluate the scaled Jacobian ($SJ$) of the resulting quad meshes and report the Hausdorff distance $d_h$. Achieving high $SJ$ values is not our primary objective, as both our curated dataset and the generated layouts permit non-orthogonal quads. We further note that the compared methods prioritize lower geometric error and may produce meshes with more than 100k faces, which can readily yield lower $d_h$ values.

\subsection{Experiment Analysis} \label{subsec:experiment}

\begin{table}[t]
    \caption{Performance evaluation of different methods on test datasets. Metrics are averaged over all test models. }
    \centering

    \sisetup{
        table-format=4.2,
        table-number-alignment = center,
    }
    \scalebox{0.9}{
        \begin{tabular}{c
                c
                S[table-format=1.2]  
                S[table-format=4.2]  
                S[table-format=4.2]  
                S[table-format=2.2]  
                S[table-format=0.2]  
            }
            \toprule
            \thead{Dataset} & \thead{Method}  $\uparrow$ & \thead{$\bm{S}_l$}  $\uparrow$ & \thead{$N_c$} $\downarrow$ & \thead{$N_I$} $\downarrow$ & \thead{$d_h$} & \thead{$SJ$}    \\
            \midrule
            \multirow{6}{*}{Part1k}
                            & \textgray{Ground-Truth}    & \textgray{1.00}                & \textgray{\quad 23.33}     & \textgray{\;\; 10.86}      & \textgray{0}  & \textgray{0.96} \\
                            & QuadriFlow                 & 0.55                           & 1645.69                    & 48.97                      & 3.38          & 0.96            \\
                            & QuadWild                   & 0.86                           & 127.21                     & \bfseries 11.50            & 2.48          & 0.95            \\
                            & QuadRemesher               & 0.82                           & 558.83                     & 12.86                      & 2.55          & 0.99            \\
                            & FSCP                       & 0.79                           & 685.15                     & 12.69                      & 2.60          & 0.98            \\
                            & \name                      & \bfseries 0.95                 & \bfseries 47.65            & 18.19                      & 7.07          & 0.93            \\ 
            \midrule
            \multirow{5}{*}{ABC1k}
                            & QuadriFlow                 & 0.26                           & 2468.44                    & \bfseries 57.54            & 2.74          & 0.98            \\
                            & QuadWild                   & 0.64                           & 5057.44                    & 75.94                      & 0.81          & 0.97            \\
                            & QuadRemesher               & 0.69                           & 2887.89                    & 61.56                      & 1.03          & 0.99            \\
                            & FSCP                       & 0.61                           & 4074.68                    & 66.44                      & 0.96          & 0.99            \\
                            & \name                      & \bfseries 0.78                 & \bfseries 216.73           & 58.37                      & 4.99          & 0.90            \\ 
            \midrule
            \multirow{5}{*}{Model300}
                            & QuadriFlow                 & 0.14                           & 3106.90                    & \bfseries 111.96           & 4.74          & 0.96            \\
                            & QuadWild                   & 0.38                           & 9032.79                    & 184.99                     & 3.20          & 0.94            \\
                            & QuadRemesher               & 0.37                           & 7637.73                    & 126.77                     & 2.87          & 0.98            \\
                            & FSCP                       & 0.31                           & 6626.65                    & 196.05                     & 2.60          & 0.96            \\
                            & \name                      & \bfseries 0.48                 & \bfseries 670.77           & 164.45                     & 16.93         & 0.84            \\ 
            \bottomrule
        \end{tabular}
    }
    \label{tab:benchmark}
\end{table}

We constructed three test datasets to evaluate \name and competing methods:
\begin{enumerate}[leftmargin=*]\setlength\itemsep{1mm}
    \item \emph{Part1k}: 1000 single-connected meshes randomly selected from our curated dataset (derived from Objaverse), with known quad layouts and high average loop simplicity scores ($\bm{S}_l = 0.99$). This dataset is excluded from training and serves to evaluate in-domain performance.
    \item \emph{ABC1k}: 1000 single-connected CAD components randomly selected from the ABC dataset~\cite{koch2019abc}. These models lack corresponding simple quad layouts and are used to test how well our model generalizes to CAD-like geometry.
    \item \emph{Model300}: 300 mechanical and organic shapes from \cite{coudert2024quad}, excluding non-manifold models. This dataset includes highly detailed shapes such as Buddha and Dragon. Our training data does not contain such complex geometries, as most samples are part-like models and our loop simplicity threshold filters out many complicated layouts. We use this dataset as a stress test to assess robustness and generalization.
\end{enumerate}

Statistics of the resulting quad layouts from different methods on these datasets are summarized in \cref{tab:benchmark}. \cref{fig:teaser} provides a gallery of our synthesized CDFs and extracted quad layouts, and \cref{fig:vis_benchmark} presents side-by-side visual comparisons on representative models.  In visualization, we assign complex charts with random colors to highlight layout simplicity: \emph{clear patterns with fewer and larger charts indicate simpler layouts}.  More visual comparisons are provided in the supplementary material. The histograms of loop simplicity scores for our results are shown in \cref{fig:histbenchmark}.

\begin{figure}[t]
    \imglabel{\columnwidth}{histbenchmarknew}{
        \put(0.08,-0.03){\small \textbf{(a) Part1k}}
        \put(0.42,-0.03){\small \textbf{(b) ABC1k}}
        \put(0.77,-0.03){\small \textbf{(c) Model300}}
    }
    \Description{Histograms of loop simplicity scores for our results. }
    \caption{Histograms of loop simplicity scores ($\bm{S}_l$) for our results. }
    \label{fig:histbenchmark}
    \vspace{-2mm}
\end{figure}
\begin{figure}[t]
    \centering
    \includegraphics[width=\columnwidth]{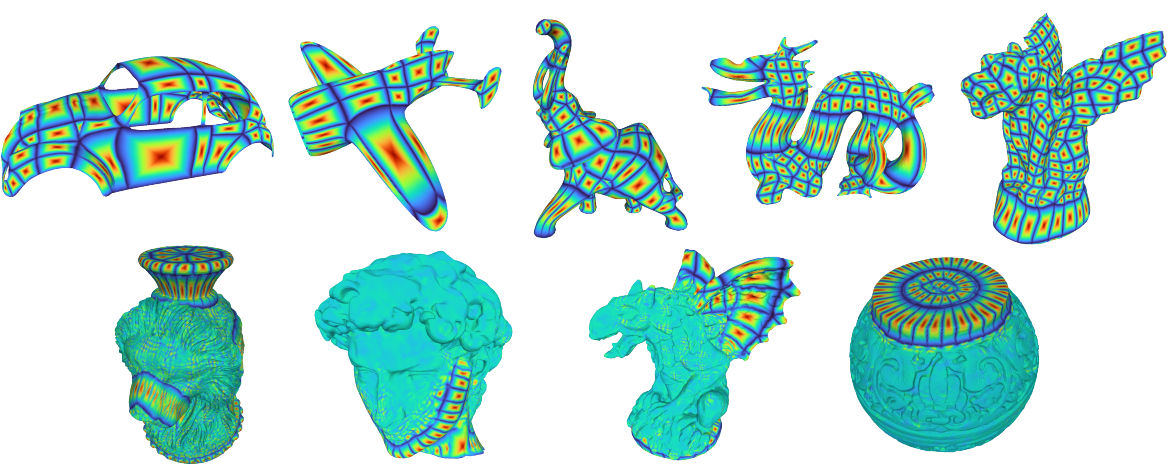}
    \Description{Generalizability and failure cases. Upper: \name produces plausible CDF patterns on complex but smooth shapes such as the Elephant and Dragon, which differ significantly from the training samples. Lower: Failure cases where \name struggles to synthesize plausible CDFs for shapes with numerous fine details.}
    \caption{Generalizability and failure cases. \textbf{Upper}: \name produces plausible CDF patterns on complex but smooth shapes such as the Elephant and Dragon, which differ significantly from the training samples. \textbf{Lower}: Failure cases where \name struggles to synthesize plausible CDFs for shapes with numerous fine details.
    }
    \label{fig:failure}
    \vspace{-4mm}
\end{figure} 
\begin{figure*}[]
    \centering
    \imglabel{.96\linewidth}{visnew}{}
\Description{Visual comparison of different methods. Examples in the three sections are selected from Part1k, ABC1k, and Model300, respectively. From left to right: input triangle meshes; quad meshes generated by QuadWild, QuadriFlow, QuadRemesher, and FSCP; and our extracted quad meshes with their synthesized CDFs. Complex charts are colored randomly to highlight layout simplicity. For some models, the synthesized CDFs contain non-quad patches and therefore require subdivision during layout extraction. Numbers shown are the loop simplicity score (blue), the complex count (orange), and the irregular vertex count (purple).}
    \caption{Visual comparison of different methods. 
Examples in the three sections are selected from Part1k, ABC1k, and Model300, respectively. 
From left to right: input triangle meshes; quad meshes generated by QuadWild~\cite{heistermann2023min}, QuadriFlow~\cite{huang2018quadriflow}, QuadRemesher~\cite{quadremesher}, and FSCP~\cite{liang2025field}; and our extracted quad meshes with their synthesized CDFs. 
Complex charts are colored randomly to highlight layout simplicity. 
For some models, the synthesized CDFs contain non-quad patches and therefore require subdivision during layout extraction. 
Numbers shown are the loop simplicity score $S_l$ (\textblue{blue}), the complex count $N_c$ (\textorange{orange}), and the irregular vertex count $N_I$ (\textpurple{purple}).
}
    \label{fig:vis_benchmark}
\end{figure*} 

\paragraph{Evaluation on ABC1k and Part1k}
On the Part1k dataset, \name achieves high loop simplicity, comparable to ground truth in majority (as seen from \cref{fig:histbenchmark}(a)). The average chart count ($N_c$) is significantly lower than other methods, indicating that \name produces much simpler layouts. Both QuadWild and QuadRemesher achieve low irregular vertex counts ($N_I$), but due to suboptimal placement, their layouts are more complex, as reflected by higher $N_c$ values. While QuadriFlow generates most complicated layouts as seen from its high $N_c$ and low $\bm{S}_l$.
The CAD models in ABC1k are more challenging than those in our training data. \name still outperforms other methods in layout simplicity, though the number of low-quality cases increases, as shown in the histogram in \cref{fig:histbenchmark}(b). Visual comparisons on representative models from these datasets (\cref{fig:vis_benchmark}-(Upper, Middle)) show that \name consistently produces layouts with simpler loops while preserving major sharp features. In contrast, other methods often generate complex layouts that are not suitable for editing.

\paragraph{Stress Test on Model300}
On this dataset, our average loop simplicity score drops to \num{0.48} due to shape complexity and out-of-distribution effects, though it still remains noticeably higher than those of other methods. \name demonstrates strong generalization to mechanical shapes and smooth organic surfaces—for example, the Fertility model—producing layouts with high simplicity (\cref{fig:vis_benchmark}-Lower). \cref{fig:failure}-Upper further shows good generalization on previously unseen shapes such as the Elephant and Dragon.
However, for shapes with numerous fine details (\cref{fig:failure}-Lower), \name struggles to synthesize plausible CDFs, as such geometries are largely absent from our training data, leading to invalid layouts and higher geometric deviation which contributes to the high $d_h$ value reported in \cref{tab:benchmark}. Notably, on smooth regions or mechanical components of these challenging shapes, \name still produces reasonable local patterns, indicating that the model has learned strong priors for such local geometries.

\begin{figure}[t]
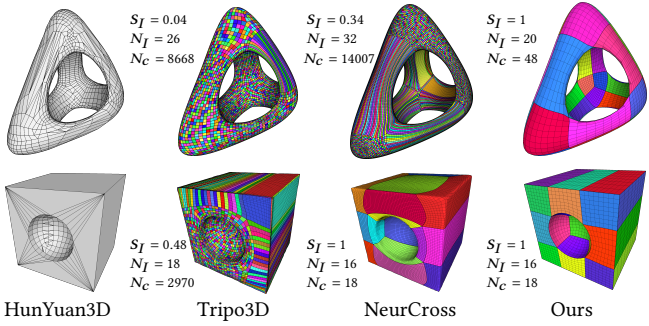

    \centering
    \imglabel{\columnwidth}{vslearnnew}{
        \put(0,-0.035){\small HunYuan3D}
        \put(0.30,-0.035){\small Tripo3D}
        \put(0.56,-0.035){\small NeurCross}
        \put(0.85,-0.035){\small Ours}

        \put(0.195,0.42){\tiny $\bm{S}_I=0.04$}
        \put(0.195,0.39){\tiny $N_I=26$}
        \put(0.195,0.36){\tiny $N_c=8668$}

        \put(0.195,0.07){\tiny $\bm{S}_I=0.48$}
        \put(0.195,0.04){\tiny $N_I=18$}
        \put(0.195,0.01){\tiny $N_c=2970$}

        \put(0.47,0.42){\tiny $\bm{S}_I=0.34$}
        \put(0.47,0.39){\tiny $N_I=32$}
        \put(0.465,0.36){\tiny $N_c=14007$}

        \put(0.47,0.07){\tiny $\bm{S}_I=1$}
        \put(0.47,0.04){\tiny $N_I=16$}
        \put(0.47,0.01){\tiny $N_c=18$}

        \put(0.75,0.42){\tiny $\bm{S}_I=1$}
        \put(0.75,0.39){\tiny $N_I=20$}
        \put(0.75,0.36){\tiny $N_c=48$}

        \put(0.75,0.07){\tiny $\bm{S}_I=1$}
        \put(0.75,0.04){\tiny $N_I=16$}
        \put(0.75,0.01){\tiny $N_c=18$}
    }
    \vspace{0.25mm}
    \Description{Comparison with learning-based methods: HunYuan3D Retopology, Tripo3D Retopology, and NeurCross. Except HunYuan3D's results, chart complexes are colorized with random colors, and layout metrics are displayed.}
    \caption{Comparison with learning-based methods: HunYuan3D Retopology~\cite{hunyuan3d}, Tripo3D Retopology~\cite{tripo3d}, and NeurCross~\cite{dong2025neurcross}. 
Except HunYuan3D's results, chart complexes are colorized with random colors, and layout metrics are displayed.}
    \label{fig:complearning}
\end{figure}

\paragraph{Comparison with Learning Approaches}
We compare \name with three implementation-accessible learning-based approaches: NeurCross~\cite{dong2025neurcross}, HunYuan3D Retopology~\cite{hunyuan3d}, and Tripo3D Retopology~\cite{tripo3d}.
NeurCross learns cross-fields via self-supervised per-shape optimization and performs quad meshing through cross-field-based parametrization.
The technical details of HunYuan3D and Tripo3D are not publicly disclosed; according to the results, HunYuan3D appears to follow an autoregressive strategy that first generates triangle meshes and then merges them into quads, while Tripo3D adopts a frame-field learning approach.
Due to NeurCross's expensive per-shape training and the latter two being accessible only through web interfaces, we evaluate them on two simple shapes that admit known simple quad layouts.
As shown in \cref{fig:complearning}, HunYuan3D produces quad-dominant meshes with many triangles and irregular vertices, and Tripo3D often introduces spiral loops, resulting in layouts with many charts. NeurCross produces smooth cross-fields but still yields unsatisfactory layouts: mismatched parameter lines lead to high chart counts in one example, and misalignment with sharp features yields poorly preserved corners in another. In contrast, \name produces clean and simple quad layouts .

\paragraph{Ablation on Latent Numbers and Regularized Inferences}
We evaluate our two proposed test-time strategies -- increasing the number of latents and applying regularized inference -- on the ABC1k dataset. As shown in \cref{tab:testtime_ablation}, increasing the number of latents from \num{4096} to \num{8192} improves layout simplicity metrics, with or without regularized inference. We also observe that using \num{8192} latents often results in larger chart counts  compared to \num{4096} latents, likely because the additional latents enable the model to capture finer geometric details, which in turn introduces more charts in the synthesized layouts. Regularized inference consistently enhances layout simplicity across all settings, regardless of the number of latents used.

\begin{figure}[t]
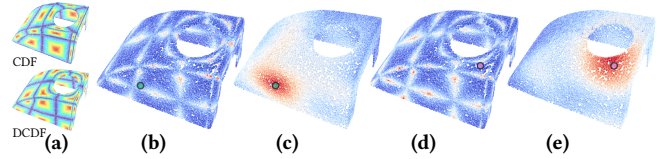

    \imglabel{\linewidth}{newvaean}{
        \put(0.00,0.10){\tiny CDF}
        \put(0.00,-0.01){\tiny DCDF}
        \put(0.05,-0.03){\small \textbf{(a)}}
        \put(0.20,-0.03){\small \textbf{(b)}}
        \put(0.41,-0.03){\small \textbf{(c)}}
        \put(0.62,-0.03){\small \textbf{(d)}}
        \put(0.83,-0.03){\small \textbf{(e)}}
    }
    \Description{Attention map visualization. (a): CDF and DCDF of the input quad mesh for SQ-VAE encoding. (b) and (d): Attention maps of SQ-VAE for two different query points (highlighted as circles). (c) and (e): Corresponding attention maps of Geom-AE for the same query points. Each attention map is normalized by its maximum value and color-coded, with red indicating higher attention and blue indicating lower attention. }
    \caption{Attention map visualization.
        \textbf{(a)}: CDF and DCDF of the input quad mesh for SQ-VAE encoding.
        \textbf{(b)} \& \textbf{(d)}: Attention maps of SQ-VAE for two different query points (highlighted as circles).
        \textbf{(c)} \& \textbf{(e)}: Corresponding attention maps of Geom-AE for the same query points.
        Each attention map is normalized by its maximum value and color-coded, with red indicating higher attention and blue indicating lower attention. }

    \label{fig:vae_attn}
\end{figure}

\paragraph{Necessity of Global Attention} Both Geom-AE and SQ-VAE inherit the global attention mechanism from 3DShape2VecSet~\cite{zhang20233dshape2vecset}. In the first cross-attention layer of the encoder, a query point (Q) attends to the features of $M$ sampled surface points (serving as \emph{Key} and \emph{Value}) to predict its output signal. Visualizing the attention weights over these $M$ points reveals a clear distinction between the two models (see \cref{fig:vae_attn}).
For Geom-AE, the attention map exhibits localized attention patterns, as the geometry feature at each query point depends primarily on nearby surface information.
However, for SQ-VAE, the attention map is highly non-local: \emph{every} sampled point contributes, and points near dual chart boundaries contribute \emph{higher} attention weights. This aligns with the nature of quad layouts --- an inherently global structure that cannot be inferred solely from local geometry. Effective learning of CDF/DCDF therefore requires global interactions among all latent variables, making the VecSet-style global attention particularly suitable for this task, albeit at higher computational cost.

\begin{table}[t]
    \caption{Ablation study of latent number and latent regularization.    }
    \centering
    \sisetup{
        table-format=4.2,
        table-number-alignment = center,
    }
    \scalebox{0.9}{
        \begin{tabular}{c
                c
                S[table-format=1.2]  
                S[table-format=4.2]  
                S[table-format=4.2]  
            }
            \toprule
            \thead{\# latents} & \thead{regularized} & \thead{$\bm{S}_l$}  $\uparrow$ & \thead{$N_c$} $\downarrow$ & \thead{$N_I$} $\downarrow$ \\
            \midrule

            4096               &                     & 0.70                           & 286.98                     & 71.23                      \\
            4096               & \cmark              & 0.77                           & \bfseries 203.63           & 59.72                      \\
            8192               &                     & 0.74                           & 371.11                     & 89.56                      \\
            8192               & \cmark              & \bfseries 0.78                 & 216.73                     & \bfseries 58.37            \\
            \bottomrule
        \end{tabular}
    }
    \label{tab:testtime_ablation}
\end{table}

\begin{figure}[t]
    \includegraphics[width=\columnwidth]{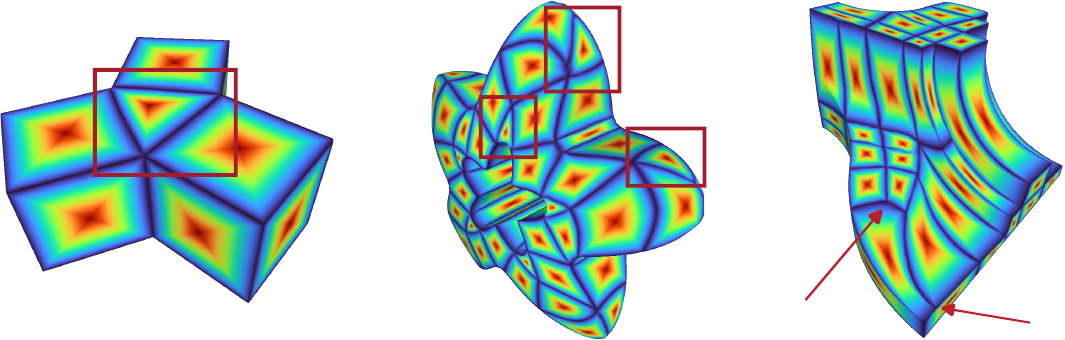}
    \Description{Examples of non-quad patches (highlighted with boxes) and T-junctions (indicated by arrows) in our results. }
    \caption{Examples of non-quad patches (highlighted with boxes) and T-junctions (indicated by arrows) in our results. }
    \label{fig:nonquad}
\end{figure}

\paragraph{Non-quad Patches and T-Junctions}
Although our training data consists of all-quad meshes, the network does not always produce perfectly quad-structured layouts. When non-quad patches appear, we apply a subdivision post-process to convert them into quadrilaterals, which typically lowers the loop simplicity scores.
However, we observe that non-quad patches and T-junctions tend to arise in \emph{meaningful} locations --- often along geometric features or in regions where they help preserve overall structural simplicity --- reflecting coherent polygonal partitioning rather than noise; even though no such quad-dominant and T-junction-containing layouts were present in the training data.
Such outputs remain practical in many modeling workflows, as quad-dominant meshes are flexible and widely used. \cref{fig:nonquad} shows representative examples.

\begin{figure}[t]
    \centering
    \includegraphics[width=\columnwidth]{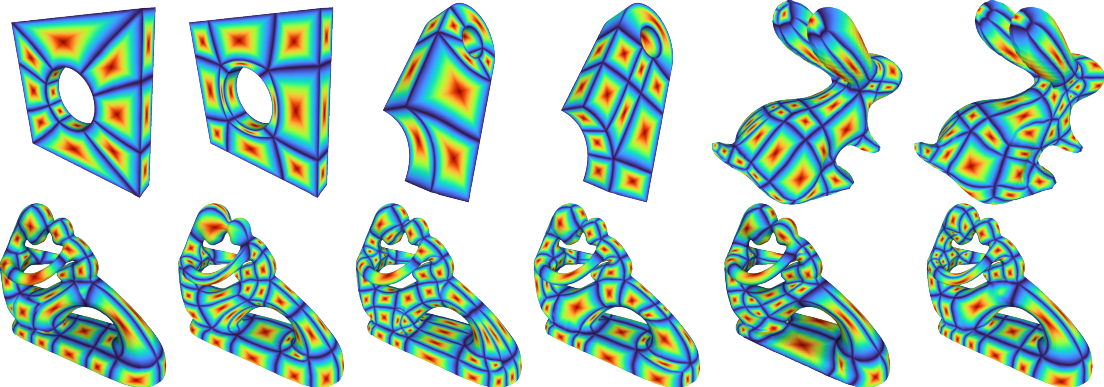}
    \Description{Generation diversity. The upper row shows two different CDFs for each input shape, obtained by sampling with different seeds, while the lower row displays six CDFs for the Fertility model.}
    \caption{Generation diversity. The upper row shows two different CDFs for each input shape, obtained by sampling with different seeds, while the lower row displays six CDFs for the Fertility model.}
    \label{fig:diversity} 
\end{figure}

\paragraph{Layout Diversity} \name can generate multiple quad layout candidates for the same input geometry. Since no formal metric is available, we evaluate layout diversity qualitatively by sampling multiple outputs for a fixed shape. \cref{fig:diversity} shows visually distinct layouts for four examples. Additional examples are provided in the supplementary material.

\section{Conclusion and Discussion} \label{sec:conclusion}

Building on the proposed CDF representation and a curated dataset of simple quad layouts, we develop a generative framework that synthesizes CDFs, which can be reliably converted into high-quality quad layouts. Extensive experiments validate both the capability of \name and the effectiveness of the CDF representation. We believe this work opens a new direction for learning-based simple layout generation, and note that CDFs/DCDFs naturally generalize to volumetric settings, suggesting potential extensions to hexahedral-dominant meshing.

The current limitations of our current approach for future improvement are summarized as follows.

\paragraph{Data Richness}
Because we focus on simple quad layouts, complex layouts and highly detailed geometries are filtered out during data preparation. Consequently, \name may not generalize well to shapes with rich geometric detail, as shown in \cref{sec:results}. Incorporating layout optimization methods to improve the simplicity of existing quad meshes and enrich the learning dataset may help mitigate this limitation.

\paragraph{Sampling Efficiency}
Due to the stochastic nature of diffusion models, different random seeds may yield varying output quality. Some generations contain noticeable artifacts, and multiple samples may be needed to obtain the best result, increasing computation cost. Exploring more efficient architectures as alternatives to the VecSet-based framework is a promising direction.

\paragraph{Layout Extraction}
Our layout extraction algorithm can misidentify chart regions when its parameters are suboptimal, resulting in less desirable quad layouts. While many such failure cases are visually easy for humans to correct, they remain difficult to resolve automatically using fixed heuristics, reflecting an inherent limitation of purely heuristic extraction strategies. This observation suggests that incorporating global layout awareness --- potentially through learning-based approaches --- could improve the robustness of layout extraction in challenging cases. At the same time, existing layout optimization methods (as reviewed in \cref{sec:relatedwork}) remain applicable and may further refine the generated layouts, and exploring their integration with our method constitutes an interesting direction.

\paragraph{Conditioning Signals}
Our method currently conditions only on geometry. Extending it to support additional conditioning signals, such as user strokes, could make the generation process more expressive and better aligned with user intent. At the same time, introducing such constraints may conflict with the existence of simple quad layouts under certain conditions, raising questions about feasibility and trade-offs. In a related vein, allowing partial quad layouts as input to guide the completion of full layouts represents another promising avenue for future exploration.

\bibliographystyle{ACM-Reference-Format}
\bibliography{src/ref}
\appendix

\section{Appendix}
\label{sec:appendix}

\subsection{Subchart coordinate computation}
\label{appendix:subchart}

Given a quad face ${\bm{q}_{00} \bm{q}_{10} \bm{q}_{11} \bm{q}_{01}}$ belongs to a subchart of a dual chart, and the vertices are associated with subchart coordinates $(a_0, b_0), (a_1, b_0), (a_1, b_1), (a_0, b_1)$, the computation of the subchart coordinate of a point $\mp$ inside the quad face is as follows.

When $\mp$ lies on the triangle $\triangle \bm{q}_{00}\bm{q}_{10}\bm{q}_{11}$, its subchart coordinate is:
\begin{align}
  p_x = a_0 + t_x (a_1 - a_0), \; p_y = b_0 + t_y (b_1 - b_0),
  \label{eq:subchartcoord1}
\end{align}
where
\begin{align}
  t_x = \frac{(\bm{\bar{p}} - \bm{\bar{q}}_{00}) \cdot (\bm{\bar{q}}_{11} - \bm{\bar{q}}_{10})^\perp}{(\bm{\bar{q}}_{10} - \bm{\bar{q}}_{00})  \cdot (\bm{\bar{q}}_{11} - \bm{\bar{q}}_{10})^\perp}, \quad
  t_y = \frac{(\bm{\bar{p}} - \bm{\bar{q}}_{10}) \cdot  (\bm{\bar{q}}_{10} - \bm{\bar{q}}_{00})^\perp}{ (\bm{\bar{q}}_{11} - \bm{\bar{q}}_{10}) \cdot (\bm{\bar{q}}_{10} - \bm{\bar{q}}_{00})^\perp}.
\end{align}
Here, assuming a rigid transformation that maps the triangle and $\mp$ to the xy-plane, the transformed versions are denoted by $\bm{\bar{p}}$, $\bm{\bar{q}}_{00}$, $\bm{\bar{q}}_{10}$, $\bm{\bar{q}}_{11} \in \mathbb{R}^2$.

Similarly, when $\mp$ lies on the triangle $\triangle \bm{q}_{00}\bm{q}_{11}\bm{q}_{01}$, its subchart coordinate is
\begin{align}
  p_x = a_0 + t_x (a_1 - a_0), \; p_y = b_0 + t_y (b_1 - b_0),
  \label{eq:subchartcoord2}
\end{align}
where
\begin{align}
  t_x = \frac{(\bm{\bar{p}} - \bm{\bar{q}}_{01}) \cdot (\bm{\bar{q}}_{01} - \bm{\bar{q}}_{00})^\perp}{(\bm{\bar{q}}_{11} - \bm{\bar{q}}_{01})  \cdot (\bm{\bar{q}}_{01} - \bm{\bar{q}}_{00})^\perp}, \quad
  t_y = \frac{(\bm{\bar{p}} - \bm{\bar{q}}_{00}) \cdot  (\bm{\bar{q}}_{11} - \bm{\bar{q}}_{01})^\perp}{ (\bm{\bar{q}}_{01} - \bm{\bar{q}}_{00}) \cdot (\bm{\bar{q}}_{11} - \bm{\bar{q}}_{01})^\perp}.
\end{align}
$\bm{\bar{p}}, \bm{\bar{q}}_{00}, \bm{\bar{q}}_{01}, \bm{\bar{q}}_{11} \in \mathbb{R}^2$ denote the transformed versions of $\mp, \bm{q}_{00}, \bm{q}_{01}, \bm{q}_{11}$, under another rigid transformation.

\subsection{CDF/DCDF Densification}
\label{appendix:densify}
\begin{figure}[t]
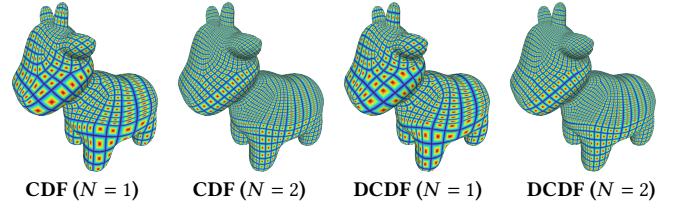

    \imglabel{\columnwidth}{densify}{
        \put (0.02,-0.038) {\small \textbf{CDF ($N=1$)}}
        \put (0.28,-0.038) {\small \textbf{CDF ($N=2$)}}
        \put (0.53,-0.038) {\small \textbf{DCDF ($N=1$)}}
        \put (0.80,-0.038) {\small \textbf{DCDF ($N=2$)}}
    }
    \vspace{0.1mm}
    \Description{Illustration of CDF and DCDF densification. }
    \caption{Illustration of CDF and DCDF densification. }
    \label{fig:densify}
\end{figure}

Since the CDF/DCDF is tightly coupled with the local parameterization (see \cref{eq:cdf,eq:dcdf,eq:para}) and is periodically distributed, we can densify the CDF/DCDF by constructing a CDF within each subchart and applying the densification recursively. This procedure is equivalent to applying the following transformation to the CDF and DCDF:
\begin{align}
  \cdf^\star(\mp)  &= 1 - 2N \cdot \max\left\{
  \left|u(\mp) - (\lfloor Nu(\mp) \rfloor + 0.5)/N \right|,\right.\nonumber\\
  &\quad\left.\left|v(\mp) - (\lfloor Nv(\mp) \rfloor + 0.5)/N \right|
  \right\}, \\
  \dcdf^\star(\mp) &= 1 - 2 \cdot \max\left\{
  \left| Nu(\mp) - \lfloor Nu(\mp) + 0.5 \rfloor \right|,\right.\nonumber\\
  &\quad\left.\left| Nv(\mp) - \lfloor Nv(\mp) + 0.5 \rfloor \right|
  \right\}.
\end{align}
Here $N \geq 1$ is the densification factor, and $\cdf^\star$ and $\dcdf^\star$ denote the densified CDF and DCDF, respectively. 
Due to the symmetry of the above equations, we can choose $(u, v) = (1-\cdf, \dcdf)$ or $(u, v) = (\dcdf, 1-\cdf)$ for the computation.
This transformation increases the number of charts by a factor of $4^N$. \cref{fig:densify} illustrates the densified CDF and DCDF for $N = 1$ and $N = 2$ on one example shown in \cref{fig:dcdfvis}.

\end{document}